\documentclass[aps,prc,showpacs,superscriptaddress,preprint]{revtex4}

\usepackage{epsfig}
\usepackage{psfrag}
\usepackage[caption=false, font=footnotesize, justification=RaggedRight]{subfig}
\usepackage[utf8]{inputenc}
\usepackage{float,amsmath,slashed,amssymb}
\usepackage{graphicx}
\usepackage{placeins}
\usepackage{multirow}

\graphicspath{{./}}
\DeclareGraphicsExtensions{.eps}

\begin{document}

\title{Energy loss in a partonic transport model including bremsstrahlung processes}

\author{Oliver Fochler}
\affiliation{Institut f\"ur Theoretische Physik, Goethe-Universit\"at Frankfurt
am Main \\
  Max-von-Laue-Stra\ss{}e~1,
  D-60438 Frankfurt am Main, Germany\vspace*{2mm}}

\author{Zhe Xu}
\affiliation{Institut f\"ur Theoretische Physik, Goethe-Universit\"at Frankfurt
am Main \\
  Max-von-Laue-Stra\ss{}e~1,
  D-60438 Frankfurt am Main, Germany\vspace*{2mm}}
\affiliation{Frankfurt Institute for Advanced Studies (FIAS)\\
  Ruth-Moufang-Stra\ss{}e~1,
  D-60438 Frankfurt am Main, Germany\vspace*{2mm}}

\author{Carsten Greiner}
\affiliation{Institut f\"ur Theoretische Physik, Goethe-Universit\"at Frankfurt
am Main \\
  Max-von-Laue-Stra\ss{}e~1,
  D-60438 Frankfurt am Main, Germany\vspace*{2mm}}

\begin{abstract}
A detailed investigation of the energy loss of gluons that traverse a thermal gluonic medium simulated within the perturbative QCD--based transport model BAMPS (a Boltzmann approach to multiparton scatterings) is presented in the first part of this work. For simplicity the medium response is neglected in these calculations. The energy loss from purely elastic interactions is compared to the case where radiative processes are consistently included based on the matrix element by Gunion and Bertsch. From this comparison gluon multiplication processes $gg \rightarrow ggg$ are found to be the dominant source of energy loss within the approach employed here.

The consequences for the quenching of gluons with high transverse momentum in fully dynamic simulations of $Au + Au$ collisions at the RHIC energy of $\sqrt{s} = 200\,\mathrm{AGeV}$ are discussed in the second major part of this work. The results for central collisions as discussed in a previous publication are revisited and first results on the nuclear modification factor $R_{AA}$ for non--central $Au + Au$ collisions are presented. They show a decreased quenching compared to central collisions while retaining the same shape. The investigation of the elliptic flow $v_{2}$ is extended up to non--thermal transverse momenta of $10\,\mathrm{GeV}$, exhibiting a maximum $v_{2}$ at roughly $4$~to~$5\,\mathrm{GeV}$ and a subsequent decrease.

Finally the sensitivity of the aforementioned results on the specific implementation of the effective modeling of the Landau--Pomeranchuk--Migdal (LPM) effect via a formation time based cut--off is explored.
\end{abstract}

\maketitle
\newpage
\section{Introduction} \label{sec:Introduction}
It has been established by experiments at the Relativistic Heavy Ion Collider (RHIC) that jets with high transverse momenta are suppressed in $Au + Au$ collisions with respect to a scaled $p + p$ reference \cite{Adler:2002xw,Adcox:2001jp}. This phenomenon of jet quenching \cite{Gyulassy:1993hr} is commonly attributed to energy loss on the partonic level as the jets produced in initial hard interactions traverse the hot medium, the quark--gluon plasma (QGP), created in the early stages of such extremely violent heavy ion collision. Due to the large momentum scales involved the energy loss of partonic jets can be treated on grounds of perturbative QCD (pQCD). Most theoretical schemes attribute the main contribution to partonic energy loss to radiative processes, where gluons are emitted in bremsstrahlungs--like interactions \cite{Zakharov:1996fv, Baier:1996sk,Baier:1998yf,Gyulassy:2000er,Jeon:2003gi,Salgado:2003gb,Wicks:2005gt}.

It is a major challenge to combine these approaches with models that describe the bulk properties of the medium, most notably the strong elliptic flow, quantified by the Fourier parameter $v_{2}$. Comparison to hydrodynamical calculations shows that the viscosity of the QGP is quite small \cite{Romatschke:2007mq}, possibly close to the conjectured lower bound $\frac{\eta}{s} = \frac{1}{4 \pi}$ from a correspondence between conformal field theory and string theory in an Anti-de-Sitter space \cite{Kovtun:2004de}. Results from hydrodynamical simulations are used as an input for the medium evolution in jet--quenching calculations (see \cite{Bass:2008rv} for an overview) and as ingredients in sophisticated Monte Carlo event generators \cite{Schenke:2009gb}. However, these approaches treat medium physics and jet physics in the QGP on very different grounds. In a previous publication \cite{Fochler:2008ts} we have established that partonic transport models including radiative pQCD interactions can provide means to investigate bulk properties of the QGP and the evolution of high--energy gluon jets within a common physical framework.

There the build up of elliptic flow and the quenching of gluon jets have been studied in fully dynamic microscopic simulations of a gluonic medium created in $Au + Au$ collisions employing mini--jet initial conditions. Matching the experimental values on integrated $v_{2}$, the nuclear modification factor $R_{AA}$ was found to be completely flat and slightly below the value obtained by Wicks et al. in their calculations for gluonic $R_{AA}$ \cite{Wicks:2005gt}. We clearly demonstrated that the incorporation of inelastic bremsstrahlung processes into partonic transport models provides a promising approach and that the consequences of such a description need to be investigated carefully.

In this work we focus on the energy loss mechanisms within our parton cascade BAMPS (a Boltzmann approach to multiparton scatterings) \cite{Xu:2004mz,Xu:2007aa} and in particular provide details on the evolution of high--energy gluons that traverse a thermal gluonic medium. We present our results in such ways that will allow other partonic transport models to easily compare with our findings \footnote{To this end we adopt the propositions for computational setups made by the TECHQM collaboration (https://wiki.bnl.gov/TECHQM), known as the ``brick problem'', to a large extend.}. Note that the term ``jet'' as used throughout this paper refers to a single parton with high energy and does not fully coincide with the experimental notion.

This paper is structured as followed: In section \ref{sec:BAMPS} we introduce the transport model BAMPS used in our investigations. We especially focus on the details concerning the inelastic gluon multiplication processes and the involved effective implementation of the Landau--Pomeranchuk--Migdal (LPM) effect. In section \ref{sec:box_22only} we discuss the evolution and energy loss of high--energy gluons in a static medium exclusively considering binary $gg \leftrightarrow gg$ interactions. After having established this baseline, in section \ref{sec:box_including23} we turn towards the evolution of gluonic jets in a static medium including inelastic $gg \leftrightarrow ggg$ processes. Afterward, in section \ref{sec:non_central_RHIC}, we extend our investigations of central Au+Au collisions at RHIC energies from a previous publication to non--central collisions with an impact parameter $b=7\,\mathrm{fm}$. In section \ref{sec:sensitivity} we investigate the sensitivity of the previously obtained results on details of the modeling of the LPM effect via a cut--off in momentum--space.

\section{The transport model BAMPS} \label{sec:BAMPS}
BAMPS \cite{Xu:2004mz,Xu:2007aa} is a microscopic transport model aimed at simulating the early stage of heavy ion collisions on the partonic level via pQCD interactions consistently including parton creation and annihilation processes. The use of microscopic transport calculations provides a realistic way of understanding the complex evolution of the QGP including the full dynamics of the system and allowing for non-thermal initial conditions.
Partons within BAMPS are treated as semi-classical and massless Boltzmann particles. At this stage the model is limited to gluonic degrees of freedom though the implementation of light quarks is underway and will be presented in an upcoming publication. The investigations presented in this work, however, are exclusively dealing with high energy gluons traversing purely gluonic media and thus $N_{f}=0$ is understood for all calculations.

The interactions between partons are implemented based on leading order pQCD matrix elements from which transition probabilities are computed. These are used to sample the interactions of particles in a stochastic manner \cite{Xu:2004mz}. The test particle method is introduced to reduce statistical fluctuations and is implemented such that the mean free path is left invariant.
For elastic interactions of gluons, $gg \leftrightarrow gg$, we use the Debye screened cross section in small angle approximation
\begin{equation} \label{eq:gg_to_gg}
 \frac{d\sigma_{gg\to gg}}{dq_{\perp}^2}
= \frac{9\pi\alpha_{s}^{2}}{(\mathbf{q}_{\perp}^2+m_D^2)^2}\,\text{.}
\end{equation}
Throughout this work we employ a fixed coupling constant $\alpha_{s} = 0.3$. The Debye screening mass is computed dynamically from the local particle distribution $f=f(p,x,t)$ via
\begin{equation}
 m_{D}^{2} = d_G \pi \alpha_s \int \frac{d^3p}{(2\pi)^3} \frac{1}{p} N_c f
\text{,}
\end{equation}
where $d_G = 16$ is the gluon degeneracy factor for $N_c = 3$.

Inelastic $gg \leftrightarrow ggg$ processes are treated via an effective matrix element based on the work by Gunion and Bertsch \cite{Gunion:1981qs}. Detailed balance between gluon multiplication and annihilation processes is ensured by the relation $\left| \mathcal{M}_{gg \rightarrow ggg} \right|^{2} = d_{G} \left| \mathcal{M}_{ggg \rightarrow gg} \right|^{2}$. For the case of bremsstrahlung-like processes the matrix element employed in BAMPS reads
\begin{equation} \label{eq:gg_to_ggg}
\left|\mathcal{M}_{gg \to ggg}\right|^2 = \frac{72 \pi^2 \alpha_s^2 s^2}{(\mathbf{q}_{\perp}^2+m_D^2)^2}\,
 \frac{48 \pi \alpha_s \mathbf{q}_{\perp}^2}{\mathbf{k}_{\perp}^2 [(\mathbf{k}_{\perp}-\mathbf{q}_{\perp})^2+m_D^2]}
\Theta\left( \Lambda_g - \tau \right)
\text{.}
\end{equation}
$\mathbf{q}_{\perp}$ and $\mathbf{k}_{\perp}$ denote the perpendicular components of the momentum transfer and of the radiated gluon momentum in the center of momentum (CM) frame of the colliding particles, respectively.

When considering bremsstrahlung processes the LPM--effect \cite{Migdal:1956tc}, a coherence effect named after Landau, Pomeranchuk and Migdal, needs to be taken into account that leads to a suppression of the emission rate for high--energy particles. Since such an interference effect cannot be incorporated directly into a semi--classical microscopic transport model such as BAMPS, we choose an effective approach by introducing the Theta function in (\ref{eq:gg_to_ggg}). This implies that the formation time $\tau$ of the emitted gluon must not exceed the mean free path of the parent gluon $\Lambda_{g}$, ensuring that successive radiative processes are independent of each other. See \cite{Zapp:2008af} for a recently developed probabilistic algorithm to include the LPM effect in Monte Carlo event generators. The transfer of this approach to a full--scale microscopic transport model, however, would pose an enormous numerical challenge.

\begin{figure}[htb]
  \psfrag{p1}{$\vec{p}_1$}
  \psfrag{p2}{$\vec{p}_2$}
  \psfrag{p11}{$\vec{p}_1^{\,\prime}$}
  \psfrag{p22}{$\vec{p}_2^{\,\prime}$}
  \psfrag{p111}{$\vec{p}_1^{\,\prime\prime}$}
  \psfrag{p222}{$\vec{p}_2^{\,\prime\prime}$}
  \psfrag{S1}{$\Sigma$}
  \psfrag{S2}{$\Sigma^\prime$}
  \psfrag{S3}{$\Sigma^{\prime\prime}$}
  \psfrag{T}{$\theta$}
  \psfrag{b}{$\vec{\beta}^{\,\prime}$}
  \psfrag{bb}{$\vec{\beta}^{\,\prime\prime}$}
  \psfrag{kt}{$k_{\perp}$}
  \begin{center}
    \includegraphics[width=14cm]{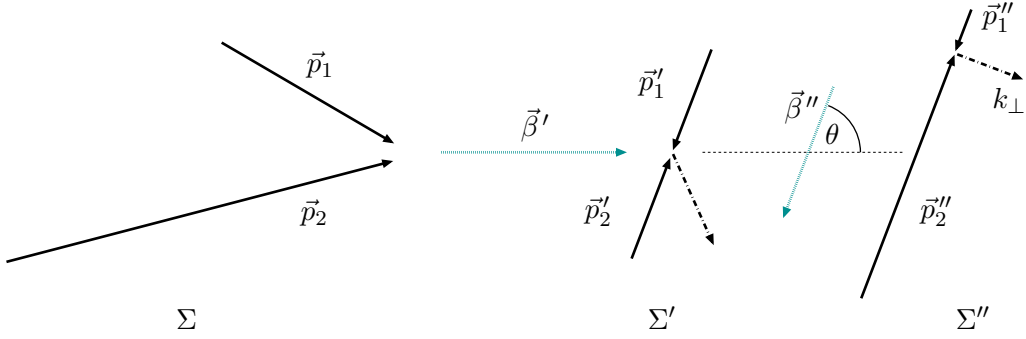}
    \caption{Illustration of the reference frames involved when comparing the mean free path $\Lambda_{g}$ measured in frame $\Sigma$ to the formation time of the emitted gluon $\tau'' = 1 / k_{\perp}$ measured in frame $\Sigma''$. $\vec{p}_i$, $\vec{p}_i^{\,\prime}$ and $\vec{p}_i^{\,\prime\prime}$ are the momenta of the incoming particles $1$ and $2$ in the respective frame, the thick dashed arrow (labeled $k_{\perp}$) depicts the radiated gluon. See text for more details.\newline
    In this example $\left| \vec{p}_2 \right| = 2 \left| \vec{p}_1 \right|$ and $\sphericalangle \left( \vec{p}_1, \vec{p}_2\right) = 45^\circ$ are chosen, leading to $\beta^{\prime}\approx 0.933$ and $\theta \approx 69^\circ$. The gluon in this example is emitted with $\cosh y =\gamma^{\prime\prime} = \sqrt{2}$.}
    \label{fig:boost_illustration} 
  \end{center}
\vspace{-0.5cm}
\end{figure}

When comparing the formation time to the mean free path, i.e. the time between successive interactions of the parent gluon, special attention needs to be paid to the frames of reference. In this case three different reference frames are involved. Let $\Sigma$ denote the local frame that is co-moving with the average velocity of the medium in each computational cell. In this frame the mean free path $\Lambda_{g}$ is computed from the interaction rates. $\Sigma'$ is the center of momentum frame of the colliding particles in which the matrix element (\ref{eq:gg_to_ggg}) is computed. Finally, $\Sigma''$ is the reference frame in which the gluon is emitted purely transversal with respect to the axis defined by the colliding particles in the CM frame and thus $\tau'' = 1 / k_{\perp}$. 

In order to compare $\Lambda_{g}$ to $\tau''$ in the Theta function modeling the LPM cut--off via $\Theta\left( \Lambda_g - \tau \right) = \Theta\left( \frac{\Lambda_g}{\gamma} - \tau'' \right)$ , the overall boost 
\begin{equation} \label{eq:gamma}
\gamma = \gamma' \gamma'' \left( 1 + \vec{\beta'}\vec{\beta''}\right) = \frac{\cosh y}{\sqrt{1-\beta'^2}}\,(1 + \beta'\, \tanh y\, \cos\theta)
\end{equation}
from $\Sigma$ to $\Sigma''$ needs to be taken into account. $\gamma'$ and $\beta'$ denote the boost and the boost velocity respectively from $\Sigma$ to $\Sigma'$. $\gamma'' = \cosh y$ and $\beta'' = \tanh y$ are the boost and boost velocity from $\Sigma'$ to $\Sigma''$. The latter can be expressed in terms of the rapidity $y$ of the emitted gluon measured from the CM frame $\Sigma'$. $\theta$ is the angle ($0 \leq \theta < \pi/2$) between $\vec{\beta'}$ and the axis of the colliding particles in the CM frame as seen from $\Sigma$. See Fig. \ref{fig:boost_illustration} for an exemplary illustration.

With this the Theta function entering the bremsstrahlung matrix element can be written as 
\begin{equation}
\label{eq:theta_function}
\Theta\left( \Lambda_g - \tau \right) = \Theta\left( k_{\perp} - \frac{\gamma}{\Lambda_g} \right) = \Theta\left( k_{\perp} \Lambda_{g} - \frac{\cosh y}{\sqrt{1-\beta'^2}}\,(1 + \beta'\, \tanh y\, \cos\theta) \right)
\text{.}
\end{equation}
For thermal energies the boost velocity $\beta'$ becomes small, $\gamma \approx \cosh y$, and the Theta function effectively reduces to $\Theta(k_{\perp}\Lambda_g - \cosh y)$ as employed in \cite{Xu:2004mz}.

\begin{figure}[p]
  \begin{center}
    \includegraphics[width=14cm]{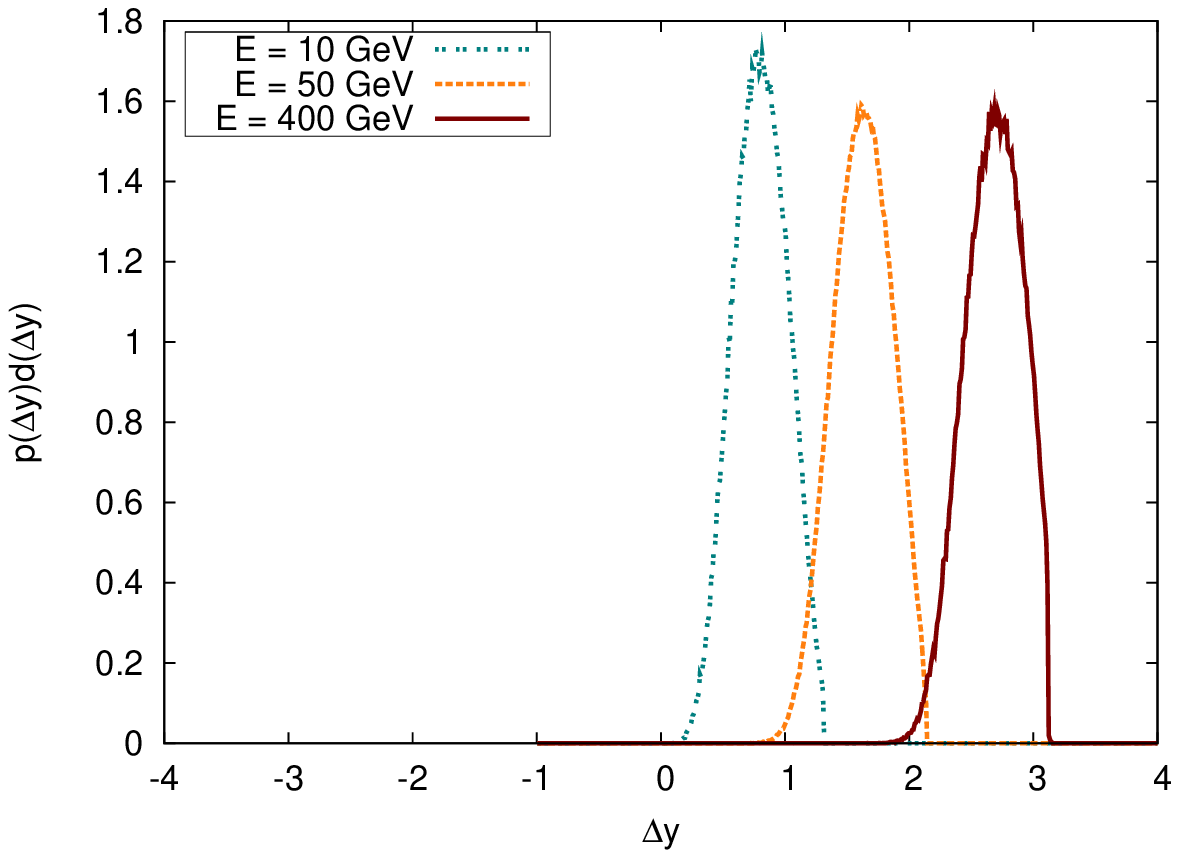}
    \includegraphics[width=14cm]{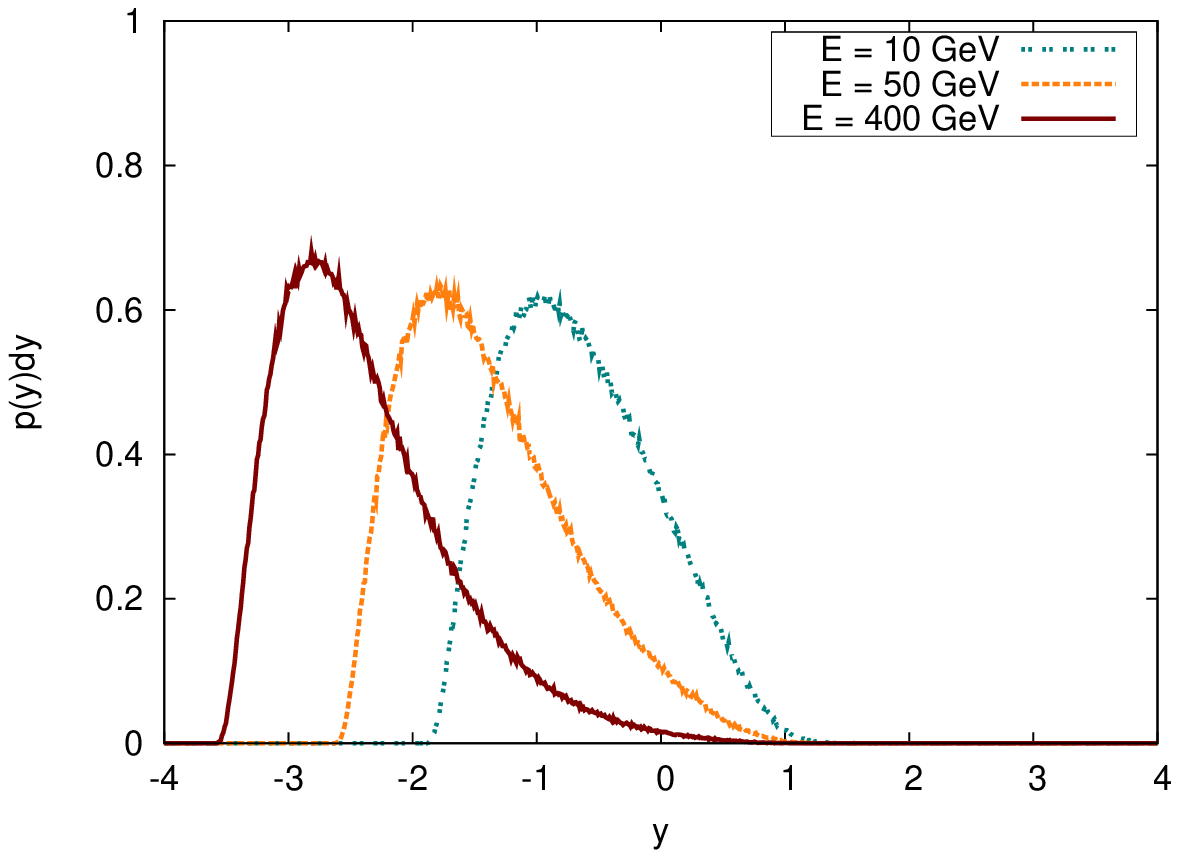}
    \caption{(Color online) Upper panel: Distribution of the magnitude of the boost from the laboratory frame to the center of momentum frame (CM) for different jet energies~$E$ (laboratory system) and $T=400\,\mathrm{MeV}$ expressed in terms of the rapidity $\Delta y = \text{arctanh}( \beta^{\prime} )$. \newline
    Lower panel: Rapidity distribution of the emitted gluon in the CM frame with respect to the incoming gluon jet momentum $\vec{p}_1^{\,\prime}$ (CM) for different jet energies~$E$ (laboratory system). The medium temperature is $T=400\,\mathrm{MeV}$.}
    \label{fig:y_distribution} 
  \end{center}
\vspace{-0.5cm}
\end{figure}

For large jet energies the boost factor (\ref{eq:gamma}) reduces the total cross section with respect to the simpler expression $\Theta(k_{\perp}\Lambda_g - \cosh y)$ since the phase space for the $k_{\perp}$ integration gets reduced. More essential for the kinematics of the outgoing particles, however, is the peculiar way in which (\ref{eq:theta_function}) distorts the shape of the available phase space. Most notably the rapidity of the emitted gluon in the center of momentum frame with respect to the incoming jet momentum $\vec{p}_1^{\,\prime}$ (CM) gets strongly shifted to negative values with increasing $\gamma$, see lower panel of Fig. \ref{fig:y_distribution}. For comparison the upper panel of Fig. \ref{fig:y_distribution} shows the rapidity $\Delta y = \text{arctanh}( \beta^{\prime} )$ associated with the boost from laboratory to CM frame. Note that $y$ and $\Delta y$ are not additive due to the angle $\theta$ in (\ref{eq:gamma}).

While for thermal energies, $\gamma \approx \cosh y$, the available phase space for the rapidity $y$ is essentially given by a $1 / \cosh y$ shape and thus symmetric around $y=0$, for larger jet energies the boost velocity $\beta'$ becomes large and the emission in the CM frame is strongly shifted to the backward direction. With this, even for small $k_{\perp}$, the energy of the emitted gluon can become large in the CM frame but will still be small in the laboratory frame due to the boost.

Note that this is still within the scope of the approximations underlying the Gunion--Bertsch matrix element \cite{Gunion:1981qs}. Besides the requirement that the transverse momenta are small, $q_{\perp}, k_{\perp} \ll \sqrt{s}$, the main approximation that leads to the result (\ref{eq:gg_to_ggg}) is that $x q_{\perp} \ll k_{\perp}$, where $x$ is the fraction of the light cone ``+''--momentum of the jet particle carried away by the radiated gluon. More specifically, in light cone coordinates, $k = \left( x \sqrt{s}, q_{\perp}^2 / (x \sqrt{s}), q_{\perp}, 0 \right)$ describes the kinematics of the radiated particle. The limit $x q_{\perp} \ll k_{\perp}$ includes $x \approx k_{\perp} / \sqrt{s}$ corresponding to emission into the central rapidity region, $y \approx 0$, that is the main interest of the authors in \cite{Gunion:1981qs}. But it also includes the $x < k_{\perp} / \sqrt{s}$ region corresponding to backward emission, $y < 0$, and thus the shift towards negative rapidities caused by our effective implementation of the LPM effect via a cut--off does not break the validity of the result by Gunion and Bertsch that is employed in (\ref{eq:gg_to_ggg}).

\section{Jets in a static medium with $gg \leftrightarrow gg$ interactions only} \label{sec:box_22only}
To gain a better understanding of the underlying mechanisms of jet quenching in heavy ion collisions, it is important to study the evolution of jets in a simplified setup. For this we track the evolution of a high--energy gluon as it propagates through a static and thermal medium of gluons. To begin with, this section focuses on BAMPS calculations exclusively incorporating binary $gg \leftrightarrow gg$ interactions in order to provide a baseline for further investigations.

The straightforward way to implement such a setup within a microscopic transport model such as BAMPS would be to populate a static box with gluons according to a thermal distribution at a given temperature $T$, then to inject a jet particle with initial energy $E$ and to track its propagation through the medium. To cut down on computation time, however, we choose a more direct, Monte Carlo--type approach. For a jet particle with given energy $E$ a certain number of collision partners is generated from a thermal distribution with temperature $T$ without actually propagating any medium constituents. This method neglects possible effects of the propagating jet on the medium. To ensure consistency, we have checked this approach against full calculations of static systems within BAMPS, where the dynamics of all particles are explicitly simulated.

The mean energy loss per unit path length $dE/dx$ is then computed as follows ($c=1$)
\begin{equation}
\frac{dE}{dx} = \frac{dE}{d(ct)} = \sum_{i} \langle \Delta E^{i} \rangle R^{i} 
\end{equation}
where $i$ denotes the interaction type ($gg \rightarrow gg$, $gg \rightarrow ggg$ and $ggg \rightarrow gg$) and $R^{i}$ is the interaction rate for process $i$. $\langle \Delta E^{i} \rangle$ is the mean energy loss in a single collision of type $i$ computed as the weighted sum
\begin{equation}
\langle \Delta E^{i} \rangle = \frac{ \sum_{j=1}^{N} \left( \Delta E^{i} \right)_{j} \tilde{P}^{i}_{j} }
{\sum_{j=1}^{N} \tilde{P}_{j} }
\text{.}
\end{equation}
The individual weighting factor $\tilde{P}^{i}_{j}$ is proportional to the interaction probability for a given process. For $gg \rightarrow gg$ and $gg \rightarrow ggg$ it is $\tilde{P}^{i}_{j} = \sigma^{i}_{j} v_{j}^{\text{rel}}$, for gluon annihilation processes $ggg \rightarrow gg$ it is $\tilde{P}^{i}_{j} = \frac{I_{32}}{E_{1} E_{2} E_{3}}$ \cite{Xu:2004mz}. Here, $\sigma^{i}$ is the cross section for a given interaction type, $v^{\text{rel}}=\frac{s}{2 E_{1} E_{2}}$ is the relative velocity of two incoming massless particles and $I_{32}$ is the phase space integral over the matrix element $\left|\mathcal{M}_{ggg \to gg}\right|^2$. For the purpose of this section only $gg \rightarrow gg$ interactions are considered. We will drop this restriction in section \ref{sec:box_including23}.

\begin{figure}[tbh]
  \begin{center}
    \includegraphics[width=14cm]{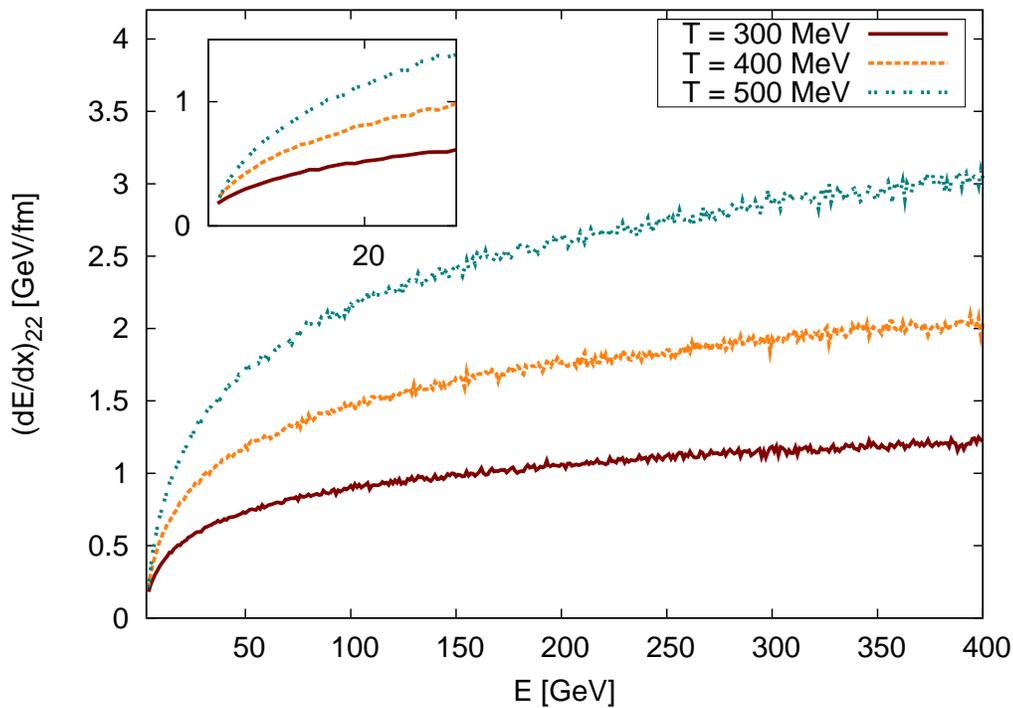}
    \caption{(Color online) Differential energy loss $(dE/dx)_{22}$ of a gluon jet in a static and thermal medium of gluons with temperatures $T=300$, $400$ and $500\,\mathrm{MeV}$ taking only binary $gg \rightarrow gg$ interactions into account.}
    \label{fig:dEdx_22_compT} 
  \end{center}
\vspace{-0.5cm}
\end{figure}

Figure \ref{fig:dEdx_22_compT} shows the mean differential energy loss $dE/dx$ of a gluon jet caused by binary $gg \rightarrow gg$ interactions as a function of the jet energy $E$ and for different medium temperatures. The energy loss is computed as described above and the medium is represented by a thermal ensemble of gluons. The jet particle is traced such that after each collision the outgoing particle with the highest energy is considered the new jet particle.

The differential energy loss exhibits the expected (see \cite{Wicks:2005gt} for a concise overview) logarithmic dependence on the jet energy $E$ and the dominant quadratic dependence on the medium temperature $T$
\begin{equation} \label{eq:dEdx22_log_behavior}
\left.\frac{dE}{dx}\right|_{2\rightarrow 2} \propto C_{R} \pi \alpha_{s}^{2} T^{2} \ln \left( \frac{4 E T}{m_{D}^{2}} \right)
\text{,}
\end{equation}
where $C_{R}$ is the quadratic Casimir of the propagating jet, $C_{R} = C_{A} = N_{c}$ for gluons. For $T = 400\,\mathrm{MeV}$ and a jet energy $E = 50\,\mathrm{GeV}$ we find an elastic energy loss of  $\left.\frac{dE}{dx}\right|_{2\rightarrow 2} = 1.2\,\mathrm{GeV}/\mathrm{fm}$ that increases to $\left.\frac{dE}{dx}\right|_{2\rightarrow 2} = 2\,\mathrm{GeV}/\mathrm{fm}$ at $E=400\,\mathrm{GeV}$.

More detailed information than in the mean energy loss per unit path length is contained in the time evolution of the energy distribution of the jet particle propagating through the medium. As for the differential energy loss a Monte Carlo approach is chosen where the collision partners are sampled from a thermal distribution. Given discretized time steps $\Delta t$, the gluon density $n(T)$ and a fixed number of collision partners $\tilde{N}$, the implemented interactions are sampled according to their probabilities
\begin{align}
P_{22} &= v_{rel} \sigma_{22} \frac{n(t) \Delta t}{\tilde{N}} & P_{23} &= v_{rel} \sigma_{23} \frac{ n(t) \Delta t}{\tilde{N}} & P_{32} &= \frac{I_{32}}{8 E_{1} E_{2} E_{3}} \frac{n(t)^{2} \Delta t}{\tilde{N}^{2}} \text{.}
\end{align}
Starting at $t=0\,\mathrm{fm/c}$ with an initial parton energy $E_{0}$ this approach yields the evolution of the energy as a function of time, $E(t)$. Repeating this procedure many times we compute $p(E)\,dE$, the probability that a parton that started with $E(t=0\,\mathrm{fm/c})=E_{0}$ has an energy $E \le E(t) < E + dE$ at a given time $t$.

\begin{figure}[p]
  \begin{center}
    \includegraphics[width=14cm]{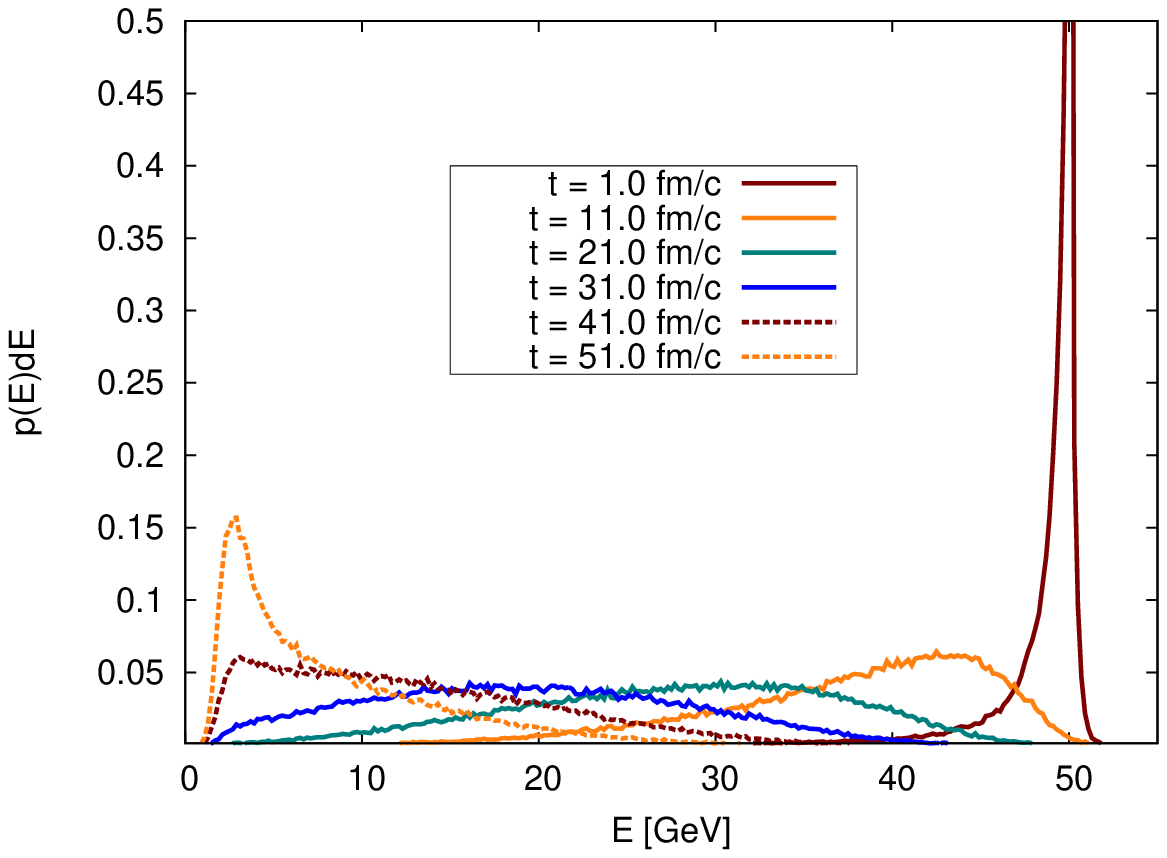}
    \includegraphics[width=14cm]{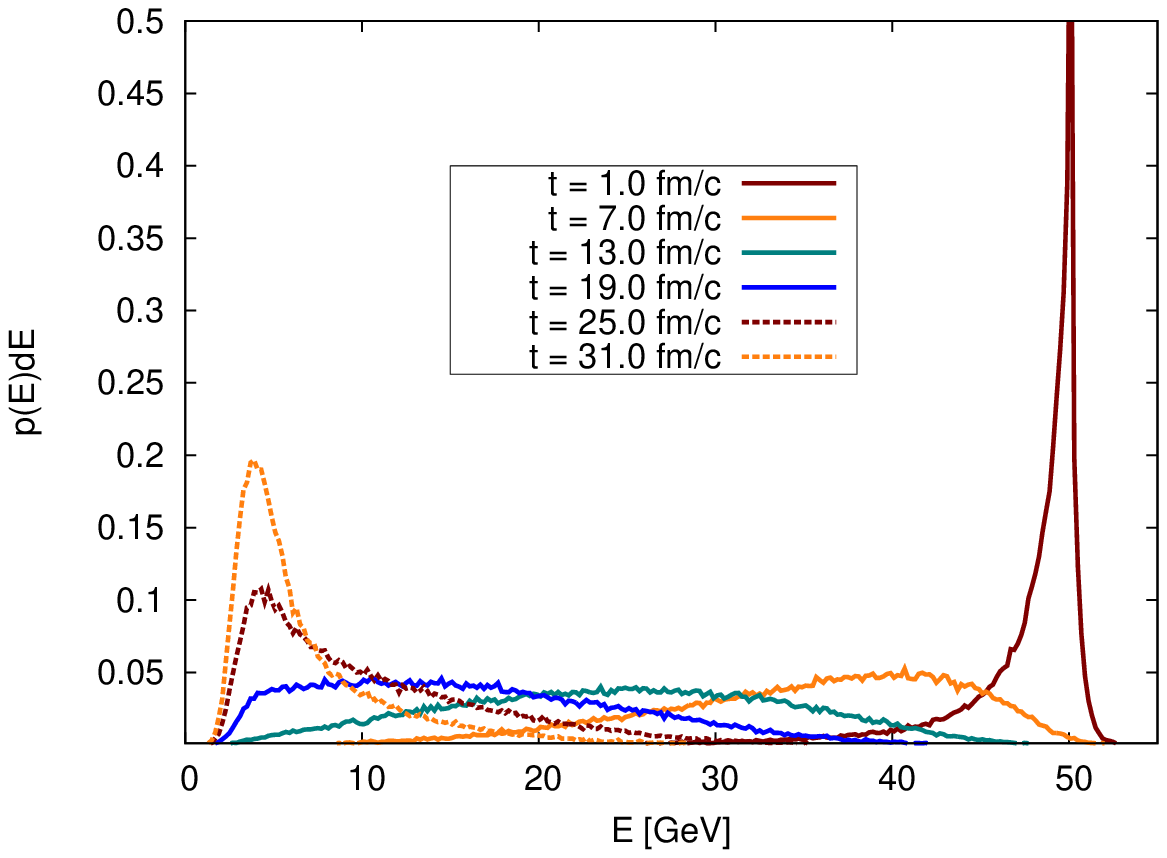}
    \caption{(Color online) Time evolution of the energy distribution of a gluon jet that interacts only via $gg \rightarrow gg$ processes with a static and thermal medium of gluons with $T=400\,\mathrm{MeV}$ (upper panel) and $T=600\,\mathrm{MeV}$ (lower panel). The initial ($t=0\,\mathrm{fm/c}$) energy of the gluon jet is $E_{0}=50\,\mathrm{GeV}$.}
  \label{fig:evolution_22_T400}
  \end{center}
\vspace{-0.5cm}
\end{figure}

Figure \ref{fig:evolution_22_T400} shows the probability distribution for the energy of a gluonic jet particle injected with an initial energy of $E_{0} = 50\,\mathrm{GeV}$ after certain propagation times in a thermal medium of $T=400\,\mathrm{MeV}$ and $T=600\,\mathrm{MeV}$ respectively. In both cases the distribution of the jet energy induced by collisions with the constituents of the medium becomes rather broad. A distinct peak at lower energies only re-emerges at very late times, roughly after $50\,\mathrm{fm/c}$ for $T=400\,\mathrm{MeV}$ and $30\,\mathrm{fm/c}$ for $T=600\,\mathrm{MeV}$. The mean energy loss as depicted in Fig. \ref{fig:dEdx_22_compT} is therefore a valuable observable but contains only limited information. It is noteworthy that there exists a finite probability for the jet to gain energy by collisions with the thermal gluons. This effect is more pronounced for higher medium temperatures.

As already found in \cite{Schenke:2009gb} the shapes of the distributions induced by collisional energy loss significantly differ from models that employ a mean energy loss accompanied by momentum diffusion such as \cite{Wicks:2005gt, Qin:2007rn}.

\section{Jets in a static medium including inelastic $gg \leftrightarrow ggg$ interactions} \label{sec:box_including23}

Having investigated the evolution of a gluon jet that exclusively undergoes elastic $gg \rightarrow gg$ interactions, we now make use of the full potential offered by BAMPS and include inelastic $gg \leftrightarrow ggg$ processes according to the matrix element (\ref{eq:gg_to_ggg}). As in the previous section we explore the propagation of gluonic jets in a static and thermal medium of gluons to systematically analyze the energy loss mechanism in simulations with the microscopic transport model BAMPS. As for the purely elastic case, the observables presented in this section are computed by means of the Monte Carlo approach presented in section \ref{sec:box_22only}.

\begin{figure}[htbp]
  \begin{center}
    \includegraphics[width=14cm]{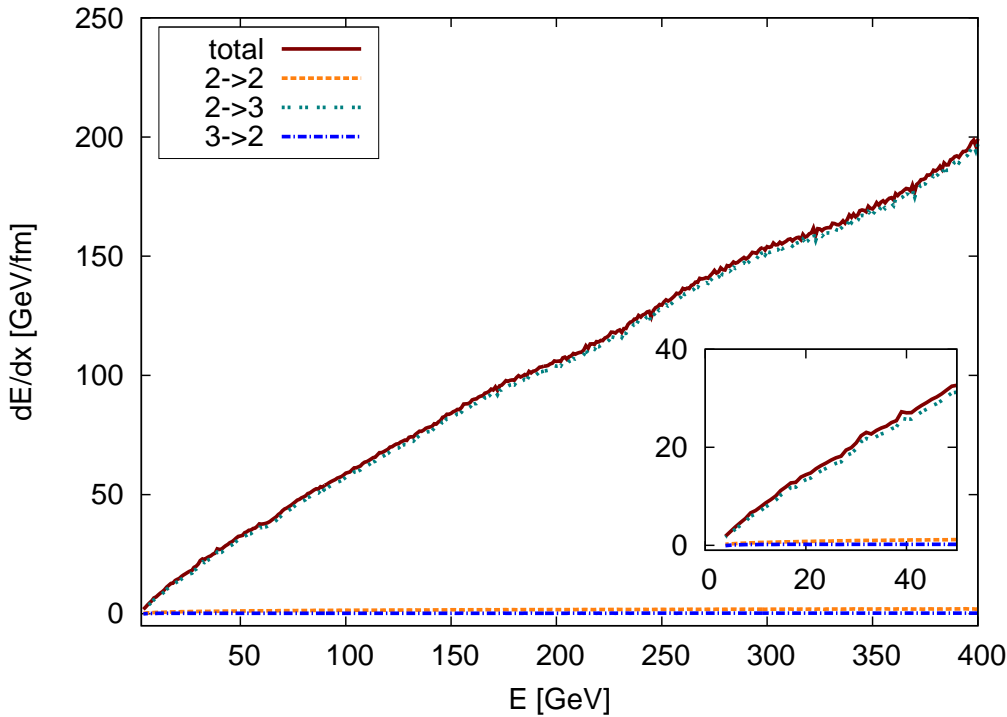}
    \caption{(Color online) Differential energy loss of a gluon jet in a static and thermal medium of gluons with $T=400\,\mathrm{MeV}$. The contributions of the different pQCD processes implemented in BAMPS ($gg \rightarrow gg$, $gg \rightarrow ggg$ and $ggg \rightarrow gg$) to the total $dE/dx$ are shown.}.\label{fig:dEdx_all_T400}
  \end{center}
\vspace{-0.5cm}
\end{figure}

Figure \ref{fig:dEdx_all_T400} shows the mean differential energy loss $dE/dx$ of a gluon jet with energy $E$ in a static thermal medium with $T=400\,\mathrm{MeV}$ caused by binary $gg \rightarrow gg$ and inelastic $gg \leftrightarrow ggg$ interactions. The contributions from the different processes to the total energy loss are displayed separately. From this compilation it is obvious that bremsstrahlung processes $gg \rightarrow ggg$ are by far the most dominant contribution to the gluonic energy loss within the BAMPS framework, whereas gluon annihilation processes are negligible and binary interactions as discussed in detail in section \ref{sec:box_22only} contribute only on a small level. The resulting differential energy loss is almost linearly rising with the energy, for example resulting in a total $dE/dx \approx 32.6\,\mathrm{GeV}/{\mathrm{fm}}$ at $E = 50\,\mathrm{GeV}$.

\begin{figure}[htbp]
  \begin{center}
    \includegraphics[width=14cm]{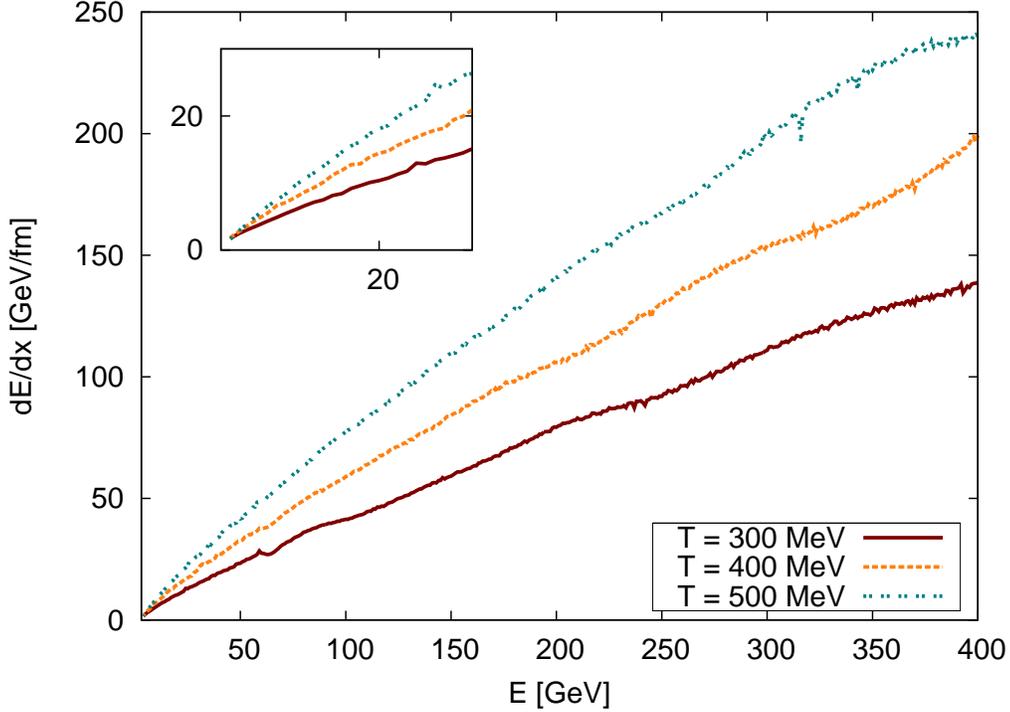}
    \caption{(Color online) Total differential energy loss, including $gg \leftrightarrow ggg$ processes, of a gluon jet in a static and thermal medium of gluons at $T=400\,\mathrm{MeV}$, $T=500\,\mathrm{MeV}$ and $T=600\,\mathrm{MeV}$.}
    \label{fig:dEdx_all_compT}
  \end{center}
\vspace{-0.5cm}
\end{figure}

At large energies the temperature dependence of the resulting total differential energy loss appears to be linear as can be seen from Fig. \ref{fig:dEdx_all_compT}, where $dE/dx$ of a gluon jet is compared for $T=400\,\mathrm{MeV}$, $T=500\,\mathrm{MeV}$ and $T=600\,\mathrm{MeV}$. This stems from the dominant $gg \rightarrow ggg$ processes and is in contrast to the elastic energy loss (\ref{eq:dEdx22_log_behavior}) that exhibits a quadratic dependence on the temperature. Possible logarithmic contributions to the temperature dependence cannot be resolved at this stage.

\begin{figure}[htbp]
  \begin{center}
    \includegraphics[width=14cm]{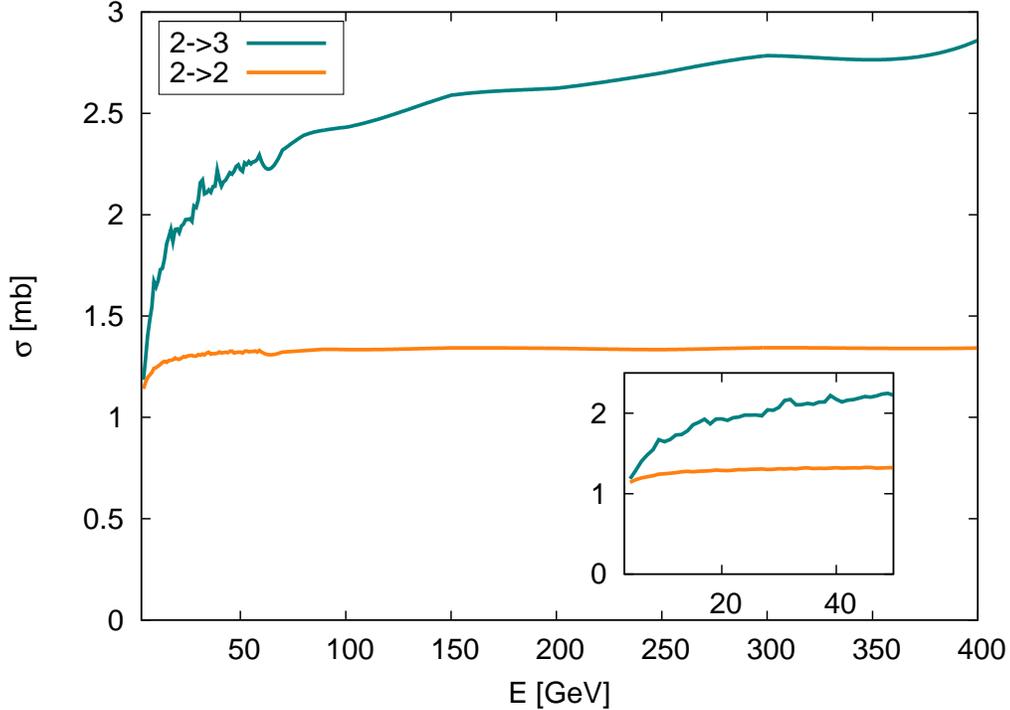}
    \caption{(Color online) Cross section of a gluon jet with energy $E$ interacting with thermal gluons ($T=400\,\mathrm{MeV}$). Shown are the cross sections for $gg \rightarrow gg$ processes and for $gg \rightarrow ggg$ processes.}
    \label{fig:sigma_T400}
  \end{center}
\vspace{-0.5cm}
\end{figure}

Given these results it is necessary to discuss the origin of the strong energy loss from radiative processes within BAMPS. First of all, despite the large differential energy loss for gluonic jets, the individual cross sections increase only slowly with the jet energy as seen in Fig. \ref{fig:sigma_T400}. For instance the average total cross sections for a gluon jet with $E=50\,\mathrm{GeV}$ in a gluonic thermal bath with a temperature $T=400\,\mathrm{MeV}$ are $\left< \sigma_{gg \rightarrow gg} \right> \approx 1.3\,\mathrm{mb}$ and $\left< \sigma_{gg \rightarrow ggg} \right> \approx 3.5\,\mathrm{mb}$. This emphasizes that BAMPS does indeed operate with reasonable partonic cross sections based on pQCD matrix elements.

\begin{figure}[htbp]
  \begin{center}
    \includegraphics[width=13cm]{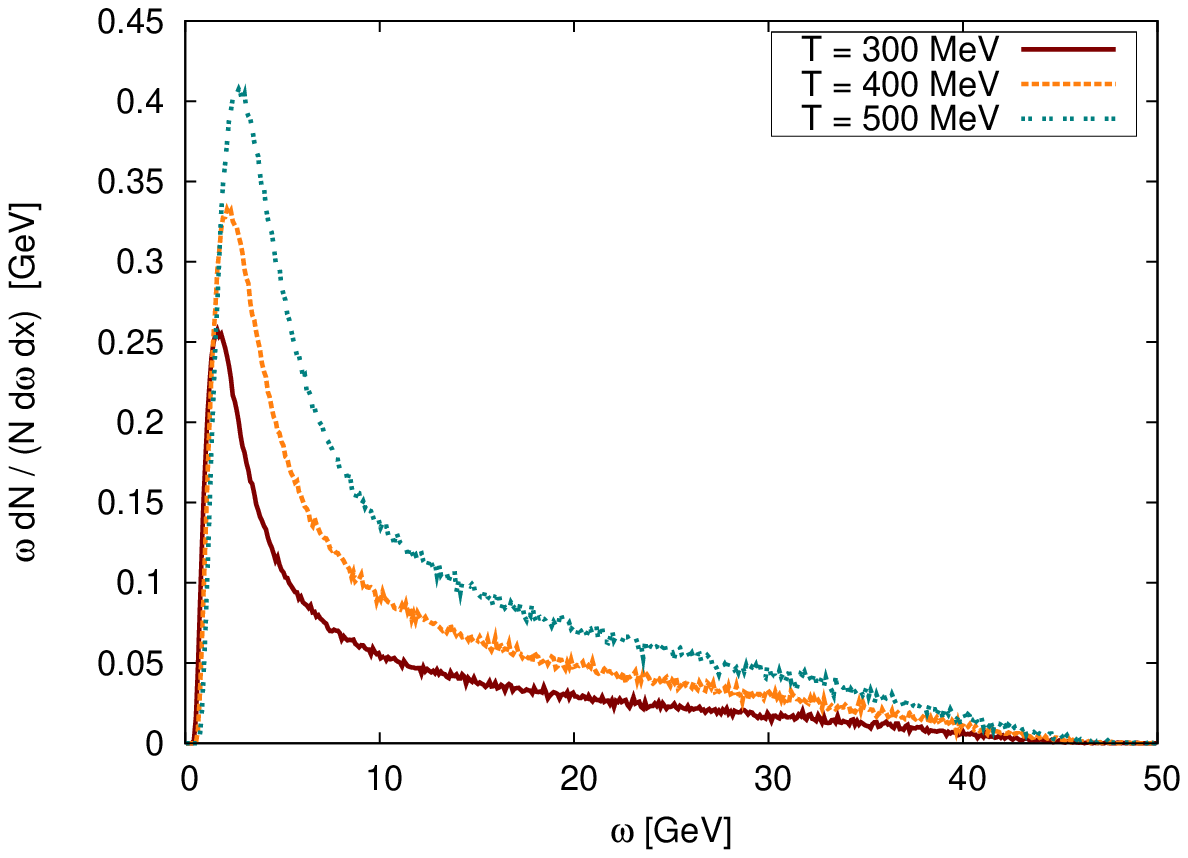}
    \includegraphics[width=13cm]{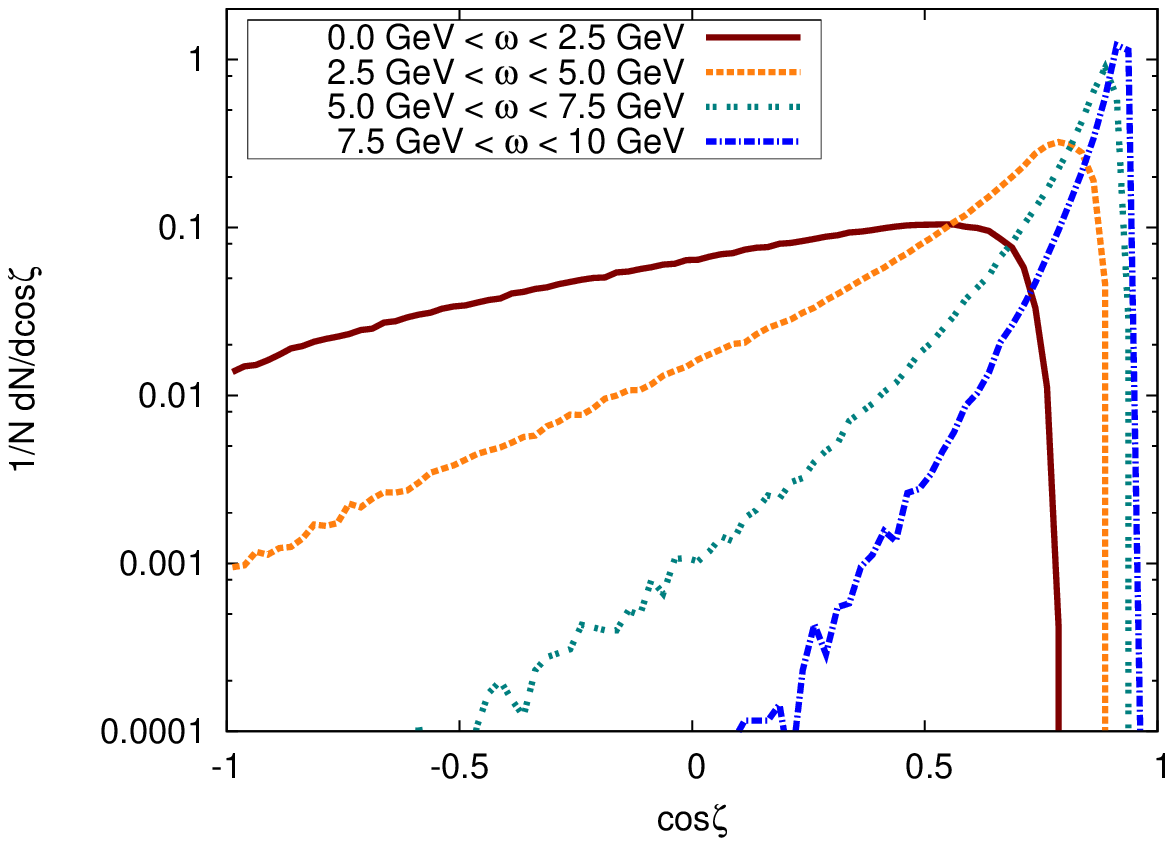}
    \caption{(Color online) Upper panel: Energy spectrum $\omega \frac{dN}{N\,d\omega\,dx}$ of radiated gluons per energy interval $d\omega$ and distance $dx$. $\omega$ is the energy (lab frame) associated with the radiated gluon according to the Gunion--Bertsch matrix element (\ref{eq:gg_to_ggg}). The radiating jet is a gluon with $E = 50\,\mathrm{GeV}$ traversing a thermal medium with $T=300\,\mathrm{MeV}$, $T=400\,\mathrm{MeV}$ or $T=500\,\mathrm{MeV}$.\newline
    Lower panel: Angular distribution of the radiated gluon in the lab frame with respect to the original jet direction for different energies $\omega$ of the radiated gluon. Jet energy $E = 50\,\mathrm{GeV}$, medium $T=400\,\mathrm{MeV}$.}
    \label{fig:rad_spectrum}
  \end{center}
\vspace{-0.5cm}
\end{figure}

Since the cross section for $gg \rightarrow ggg$ processes as seen in Fig. \ref{fig:sigma_T400} yields moderate mean free paths for jet-like particles, the cause for the large differential energy loss needs to be a large mean energy loss per single collision, $\langle \Delta E \rangle$. The energy carried away by the radiated gluons, however, is in itself not sufficient to explain such large mean $\Delta E$. From the radiation spectrum of a gluon jet with $E=50\,\mathrm{GeV}$, shown in the upper panel of Fig \ref{fig:rad_spectrum} for different medium temperatures, a mean energy of the radiated gluon can be read off that is below the $\langle \Delta E \rangle$ that would be needed to fully explain the magnitude of $dE/dx$, a finding we will explicitly confirm later. The spectrum is displayed as the number of radiated gluons per energy interval $d\omega$ and per distance $dx$ scaled by the total number of emitted gluons and weighted with the gluon energy. Here $\omega$ is the lab frame energy of the gluon that in the center of momentum frame is emitted with transverse momentum $k_{\perp}$ according to the Gunion--Bertsch matrix element (\ref{eq:gg_to_ggg}). The spectra are clearly peaked at energies that are small compared to the energy of the parent jet, with a small tail reaching out to high energies. With increasing temperature the peak of the spectrum shifts towards higher energies in an apparently linear way, favoring the emission of gluons with higher energies.

For completeness, the lower panel of Fig. \ref{fig:rad_spectrum} shows the angular distribution of gluons radiated off a $E=50\,\mathrm{GeV}$ gluon jet in a $T = 400\,\mathrm{GeV}$ medium for different ranges in the energy of the radiated gluon $\omega$. The angle $\zeta$ is taken in the lab frame with respect to the initial direction of the parent jet. With increasing energy $\omega$ the radiated gluons are emitted more preferably at small angles, only for soft gluons there is a sizable probability to be emitted transversely or in the backward direction. However, as is clearly visible in Fig. \ref{fig:rad_spectrum}, due to the cut--off in transverse momentum (\ref{eq:theta_function}) the gluons cannot be emitted at very forward angles, an effect that is more pronounced for low $\omega$.

\begin{figure}[phtb]
  \begin{center}
    \includegraphics[width=14cm]{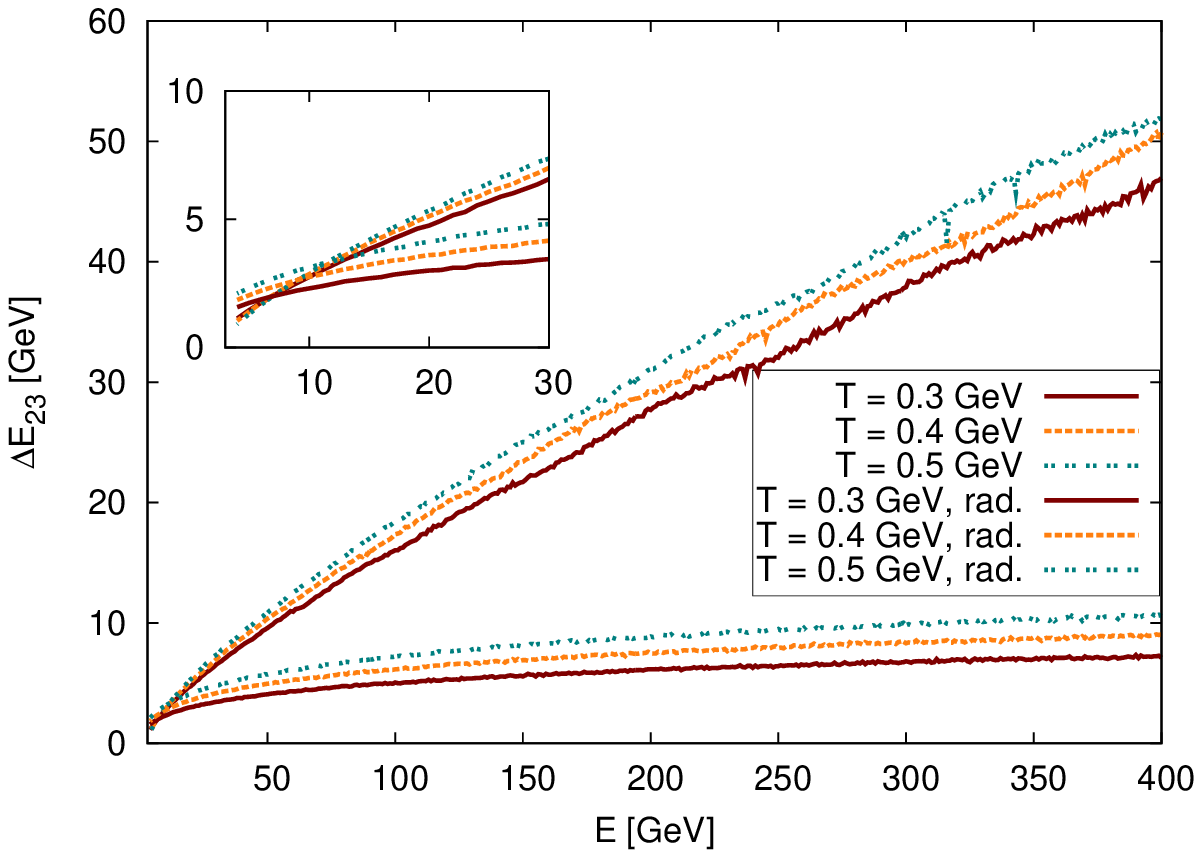}
    \includegraphics[width=14cm]{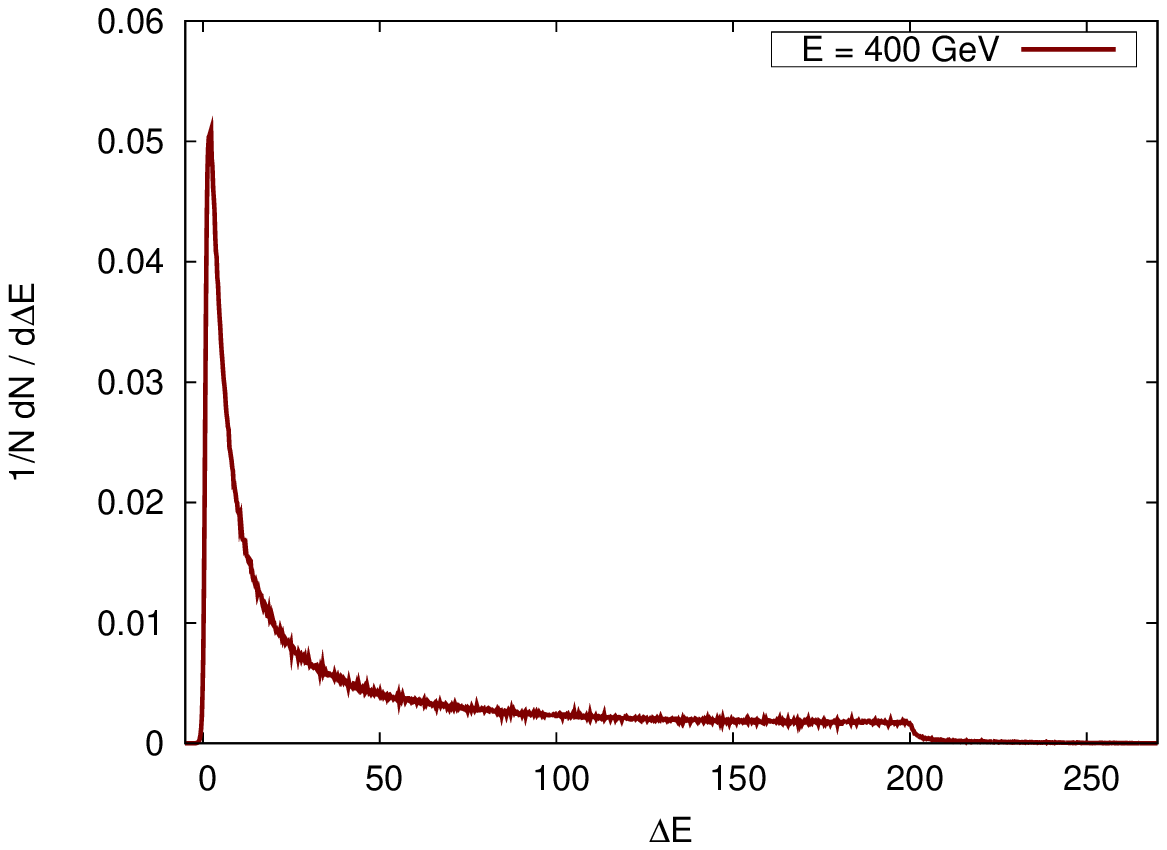}
    \caption{(Color online) Upper panel: Energy loss $\Delta E_{23}$ of a gluon jet in a single $gg \rightarrow ggg$ process. Two different cases are shown: One where $\Delta E = E^{\text{in}} - \max \left( E_{1}^{\text{out}}, E_{2}^{\text{out}}, E_{3}^{\text{out}} \right)$ (the upper group of lines) and one (labeled ``rad.'', the lower group of lines) where $\Delta E = \omega$ is the energy of the radiated gluon.\newline
    Lower panel: Distribution of $\Delta E_{23}$, with $\Delta E = E^{\text{in}} - \max \left( E_{1}^{\text{out}}, E_{2}^{\text{out}}, E_{3}^{\text{out}} \right)$, for $E = 400\,\mathrm{GeV}$ and $T=0.4\,\mathrm{GeV}$.}
    \label{fig:deltaE_details}
  \end{center}
\vspace{-0.5cm}
\end{figure}

To emphasize that the energy carried away by the radiated gluon alone does not explain the observed mean energy loss per collisions, it is noteworthy that the strong and linear rise in the energy loss due to $gg \rightarrow ggg$ is only present when identifying the outgoing particle with the highest energy as the outgoing jet and thus using $\Delta E = E^{\text{in}} - \max \left( E_{1}^{\text{out}}, E_{2}^{\text{out}}, E_{3}^{\text{out}} \right)$. This is the most natural choice and is employed throughout all calculations in this work. The average energy $\omega$ of the radiated gluon, however, is rising much slower with the jet energy, as can be seen in the upper panel of Fig. \ref{fig:deltaE_details}. This is due to the fact that the energy is distributed among three outgoing particles, the gluon emitted with energy $\omega$ being only one of them. In fact, assuming $\omega < E^{in}/3$ (in agreement with the results presented in Fig. \ref{fig:deltaE_details}), $\Delta E_{\text{min}} = \omega$ is only the smallest possible energy loss, while the largest energy loss allowed by energy and momentum conservation is $\Delta E_{\text{max}} = E^{\text{in}} - \left( \frac{E^{\text{in}} - \omega}{2} \right) = \frac{E^{\text{in}} + \omega}{2}$.

Looking at the distribution of $\Delta E$ in the lower panel of Fig. \ref{fig:deltaE_details} for a fixed jet energy that underlies the mean energy loss per collision as shown in the upper panel of Fig. \ref{fig:deltaE_details}, it becomes obvious that it is indeed a very fat tail of the $\Delta E$ distribution that causes a large averaged $\langle \Delta E \rangle$. The distinct peak at low $\Delta E$ can be readily identified as being related to energy carried away by the radiated gluon.

The reason for the heavy tail in the $\Delta E$ distribution needs to be looked for in the complicated plethora of configurations for the outgoing particles allowed by the underlying matrix element (\ref{eq:gg_to_ggg}). In the appendix, figures \ref{fig:illustrateVectors_CMS1} and \ref{fig:illustrateVectors_CMS2} show some examples for such configurations that were randomly chosen according to the matrix element (\ref{eq:gg_to_ggg}). The most specific feature of the phase space sampled in $gg \rightarrow ggg$ is that the radiated gluon is predominantly emitted into the backward hemisphere in the center of momentum frame, compare (\ref{eq:theta_function}) and Fig. \ref{fig:y_distribution}. Due to the strong bias towards negative rapidities that is present for large boosts (\ref{eq:gamma}), the energy of the radiated gluons in the CM frame is in many cases comparable to the energies of the two other outgoing particles even for small transverse momenta $k_{\perp}$.

With this some typical scenarios can be identified. Fig. \ref{fig:E3_vs_E1_qt_cut} in the appendix \ref{sec:app_typical_configurations} illustrates that for a given cut on the momentum transfer $q_{\perp}$ there are distinct configurations correlating the energy of the radiated gluon in the center of momentum frame with the energy of one of the remaining outgoing particles. Note that due to the nature of the Gunion-Bertsch matrix element (\ref{eq:gg_to_ggg}) the transverse momentum $k_{\perp}$ is typically comparable to the momentum transfer $q_{\perp}$. For example, selecting on configurations where $q_{\perp}$ and $E_{1}^{\prime}$ are small and $E_{3}^{\prime}$, i.e. the energy of the radiated gluon in the CM frame, is large, yields an energy loss significantly above its most probable value, thus these configurations contribute to the heavy tail observed in Fig. \ref{fig:deltaE_details}. This is especially true for cases where particle $1$ is emitted into the forward hemisphere, because for these specific configurations the outgoing particle $2$ needs to be emitted into the forward hemisphere as well in order to guarantee energy and momentum conservation. Boosted back into the lab frame the available energy is thus mainly split between the particles $1$ and $2$ yielding a large energy loss. The same line of reasoning holds for cases where $E_{1}^{\prime}$ is on the order of $E_{3}^{\prime}$, see example Fig. \ref{fig:illustrateVectors_CMS:d} in the appendix \ref{sec:app_examples}.

This illustrates that the fat tail in the $\Delta E$ distribution is mainly caused by configurations where in the center of momentum frame the radiated gluon is emitted with a large energy into the backward hemisphere and the remaining energy is split among the two other particles going into the forward hemisphere. Events with large $q_{\perp}$ and $k_{\perp}$ (see Fig. \ref{fig:illustrateVectors_CMS:b} and Fig. \ref{fig:illustrateVectors_CMS:e} in the appendix for examples) also yield a large energy loss but are significantly less probable due to the steeply falling $1/q_{\perp}^{4}$ contribution in the matrix element (\ref{eq:gg_to_ggg}). Compare appendix \ref{sec:app_typical_configurations} and especially Table \ref{tab:selected_regions} for a more quantitative analysis of the above mentioned configurations.

\begin{figure}[phtb]
  \begin{center}
    \includegraphics[width=14cm]{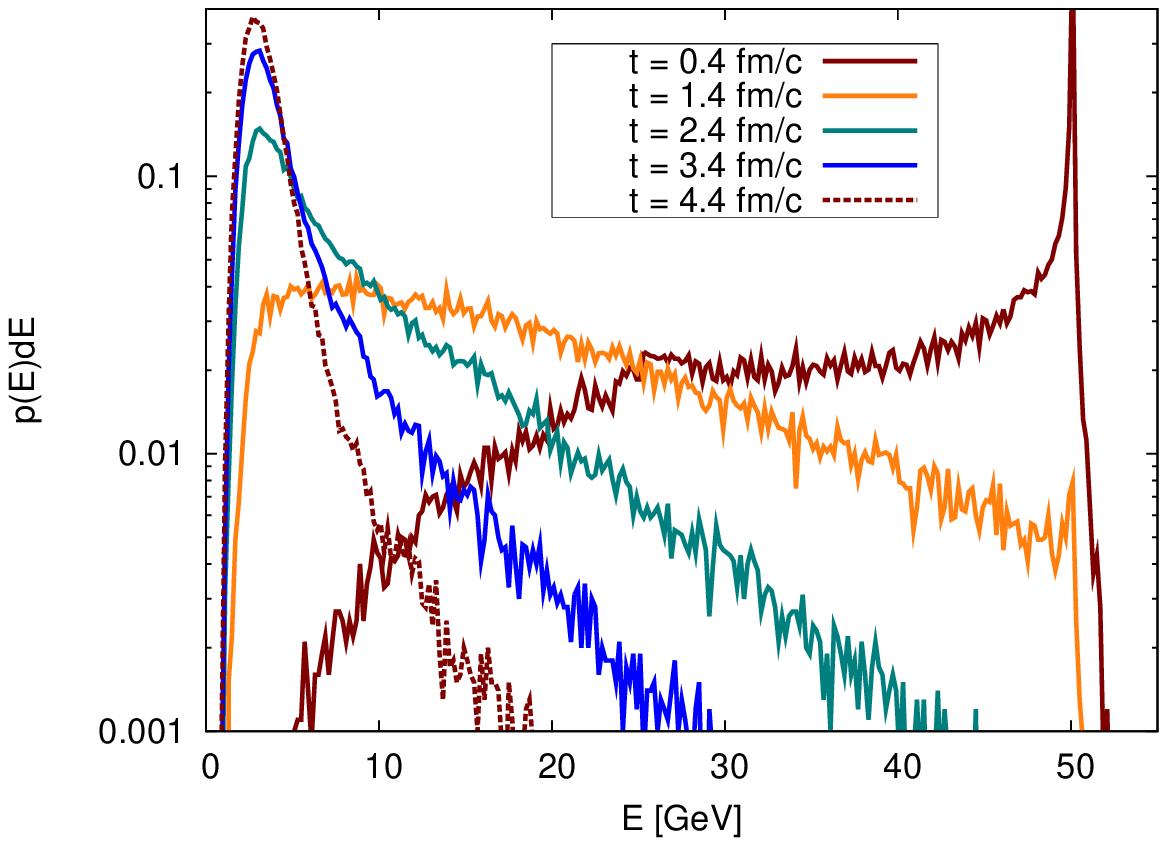}
    \includegraphics[width=14cm]{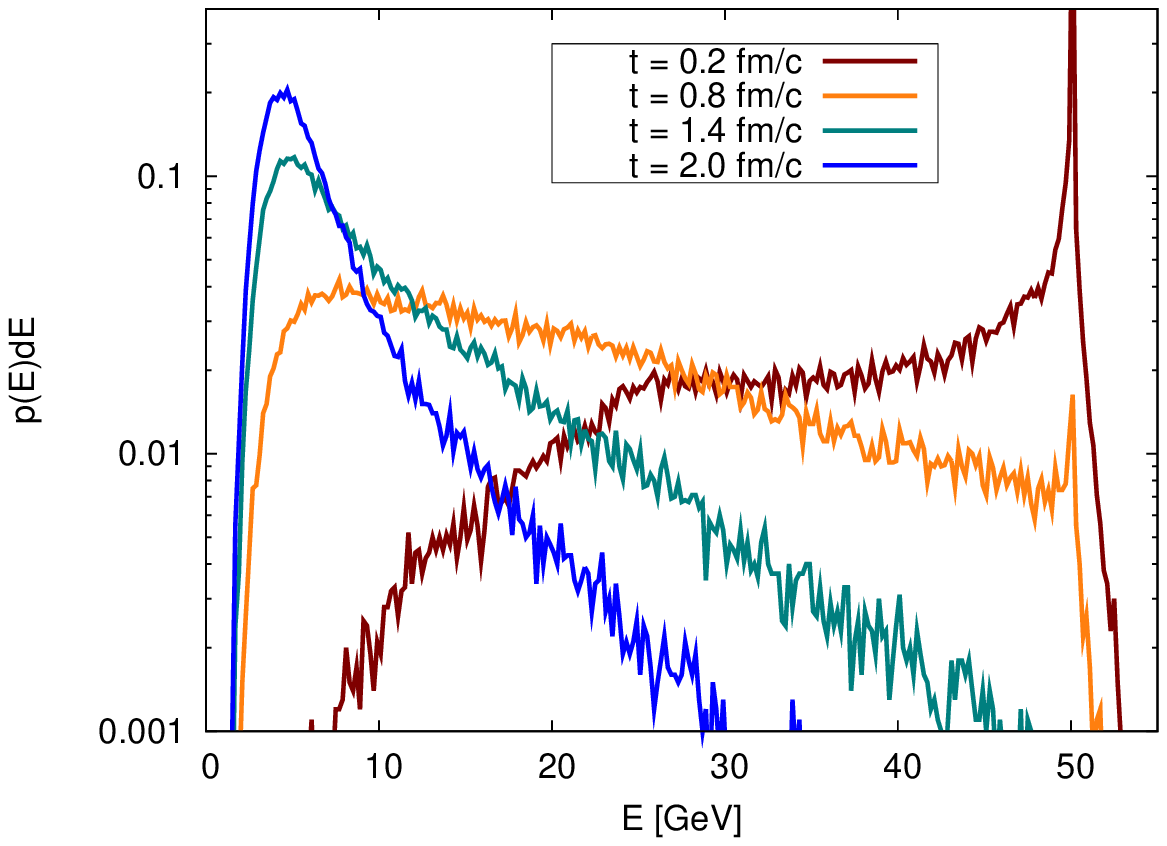}
    \caption{(Color online) Time evolution of the energy distribution of a gluon jet in a static and thermal medium of gluons with $T=400\,\mathrm{MeV}$ (upper panel) and $T=600\,\mathrm{MeV}$ (lower panel). The initial ($t=0\,\mathrm{fm/c}$) energy of the gluon jet is $E_{0}=50\,\mathrm{GeV}$. Inelastic $gg \leftrightarrow ggg$ interactions are taken into account.}
  \label{fig:evolution_all_T400}
  \end{center}
\vspace{-0.5cm}
\end{figure}

As already discussed in connection with the scenario containing only elastic interactions, valuable information beyond the mean energy loss is contained in the evolution of the energy distribution $p(E)\,dE$ of the jet. Starting out with $p(E) = \delta\left(E-E_{0} \right)$ we have seen in Fig. \ref{fig:evolution_22_T400} that elastic collisions cause a broadening of the distribution with a distinct peak at low energies only re-emerging at very large times. Because of the much stronger mean energy loss caused by inelastic $gg \rightarrow ggg$ processes a more rapid evolution is to be expected when all interactions included in BAMPS are taken into account. Indeed, Fig. \ref{fig:evolution_all_T400} shows that the energy distribution of a jet with $E_{0}=50\,\mathrm{GeV}$ traversing a gluonic medium with $T=400\,\mathrm{MeV}$ is spread over almost the entire range after roughly $1\,\mathrm{fm/c}$. A distinct peak at $E \approx \, 7.5 T$ emerges at about $3.5\,\mathrm{fm/c}$ for $T=400\,\mathrm{MeV}$ and $2\,\mathrm{fm/c}$ for $T=600\,\mathrm{MeV}$.

A quantity that is often used to characterize the effect of the medium on a jet--like particle is $\hat{q}$. It is defined as the sum of the transverse momentum transfers squared divided by the path length $L$ the particle has traveled
\begin{equation} \label{eq:qhat_def}
\hat{q} \left( L \right) = \frac{1}{L} \sum_{i} \left( \Delta p_{\perp}^{2} \right)_{i}
\text{,}
\end{equation}
where $i$ runs over all collisions the particle has undergone within the path length $L$. From now on we will simply use the time $t$ instead of $L$ since we are dealing with massless particles. Alternatively, if one knows the average momentum transfer squared per mean free path as a function of the jet energy one can compute the mean $\hat{q}$ as
\begin{equation} \label{eq:qhat_from_integral}
\langle \hat{q} \rangle \left( t \right) = \frac{1}{t} \int_{0}^{t} \left. \frac{\langle \Delta p_{\perp}^{2} \rangle}{\lambda} \right|_{E(\tilde{t})} d\tilde{t}
\text{.}
\end{equation}
Typically $\hat{q}$ is used to quantify the transverse momentum picked up from elastic collisions that eventually induce the radiation of Bremsstrahlung gluons. In the commonly used eikonal approximation the jet particle acquires no additional transverse momentum due to the radiation of the Bremsstrahlung gluons. In our approach however, radiative and elastic interactions are treated on equal grounds and jets can also pick up transverse momentum in inelastic $gg \rightarrow ggg$ processes. In the following, we therefore naturally extend the definition of $\hat{q}$ as given above to also describe the evolution of transverse momentum due to inelastic gluon multiplication processes within BAMPS.

\begin{figure}[phtb]
  \begin{center}
    \includegraphics[width=14cm]{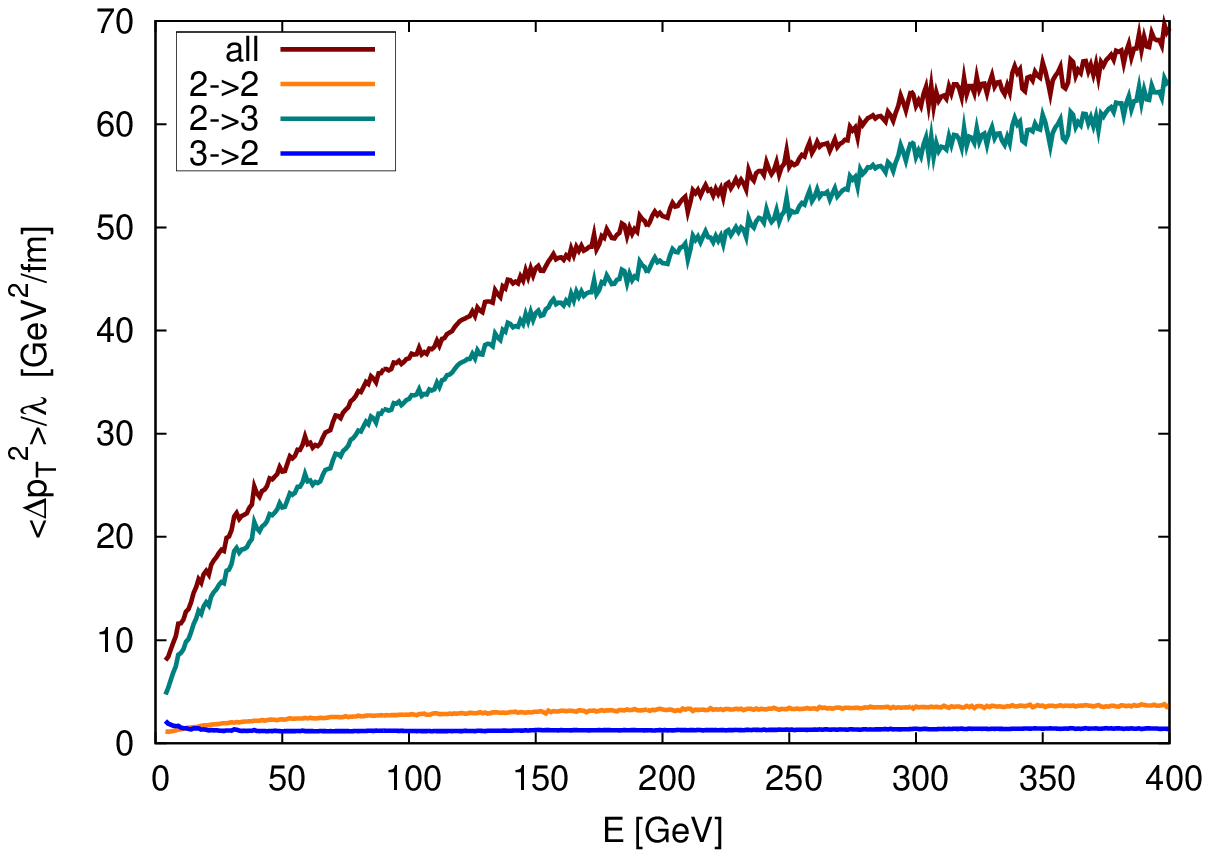}
    \includegraphics[width=14cm]{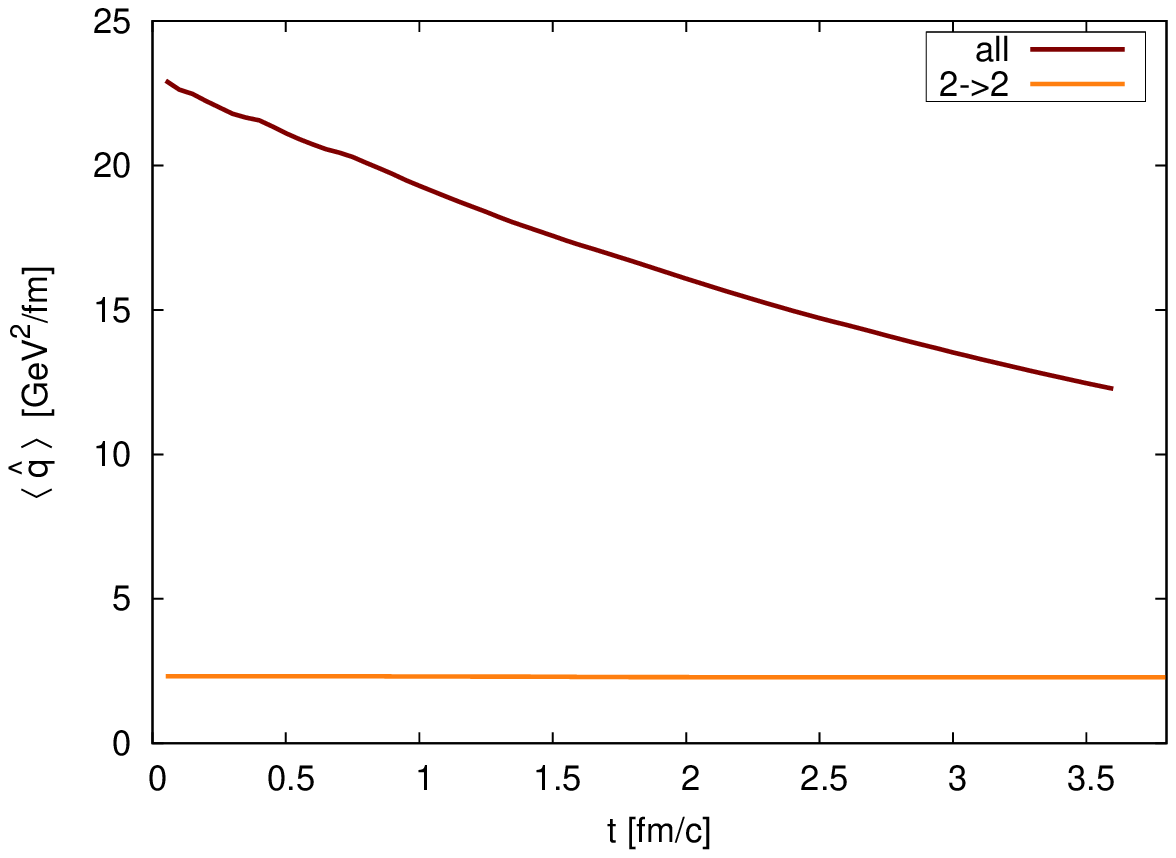}
    \caption{(Color online) Upper panel: Average transverse momentum transfer squared per mean free path, $\langle \Delta p_{\perp}^{2} \rangle / \lambda$, for $gg \rightarrow gg$, $gg \rightarrow ggg$, $ggg \rightarrow gg$ and for the sum of all processes. The medium temperature is $T=400\,\mathrm{MeV}$.\newline
    Lower panel: $\langle \hat{q} \rangle \left( t \right)$ as defined in (\ref{eq:qhat_def}) and (\ref{eq:qhat_from_integral}) for $gg \rightarrow gg$ processes and for all BAMPS processes                                                                                                                   as a function of the path length $L = t$ for a gluon jet with initial energy $E_{0}=50\,\mathrm{GeV}$. The jet traverses a medium as characterized above.}
    \label{fig:qhat_T400}
  \end{center}
\vspace{-0.5cm}
\end{figure}

As a cross--check we have compared the result from both approaches, (\ref{eq:qhat_def}) and (\ref{eq:qhat_from_integral}), using independent calculations and found perfect agreement. The upper panel of Fig. \ref{fig:qhat_T400} shows the average momentum transfer squared per mean free path $\langle \Delta p_{\perp}^{2} \rangle / \lambda$ as a function of the jet energy in a gluonic medium with $T=400\,\mathrm{MeV}$. A logarithmic behavior at large energies can be seen for $\langle \Delta p_{\perp}^{2} \rangle / \lambda$ from binary $gg \rightarrow gg$ interactions, with $\langle \Delta p_{\perp}^{2} \rangle / \lambda \approx 2.3\,\mathrm{GeV}^{2} / \mathrm{fm}$ at $E=50\,\mathrm{GeV}$ rising to $\langle \Delta p_{\perp}^{2} \rangle / \lambda \approx 3.7\,\mathrm{GeV}^{2} / \mathrm{fm}$ at $E=400\,\mathrm{GeV}$. As reflected in the differential energy loss, the average transverse momentum transfer squared per mean free path for inelastic $gg \rightarrow ggg$ interactions is much higher, $\langle \Delta p_{\perp}^{2} \rangle / \lambda \approx 22.8\,\mathrm{GeV}^{2} / \mathrm{fm}$ at $E=50\,\mathrm{GeV}$ and $\langle \Delta p_{\perp}^{2} \rangle / \lambda \approx 64.2\,\mathrm{GeV}^{2} / \mathrm{fm}$ at $E=400\,\mathrm{GeV}$ while the gluon annihilation processes $ggg \rightarrow gg$ virtually do not contribute at all.

The lower panel of Fig. \ref{fig:qhat_T400} shows $\langle \hat{q} \rangle$ as defined in equations (\ref{eq:qhat_def}) and (\ref{eq:qhat_from_integral}) as a function of the path length $L = t$ for a gluon jet with initial energy $E_{0}=50\,\mathrm{GeV}$. As before the medium is characterized by $T=400\,\mathrm{MeV}$. Over the range up to $t = 3.5\,\mathrm{fm}/\mathrm{c}$ shown in Fig. \ref{fig:qhat_T400}, the contribution from elastic interactions is almost constant at $\langle \hat{q} \rangle_{22} \approx 2.3\, \mathrm{GeV}^{2}/\mathrm{fm}$. For jets that interact only via binary $gg \rightarrow gg$ one finds that $\langle \hat{q} \rangle_{22}$ is actually slowly and linearly falling to $\langle \hat{q} \rangle_{22} \approx 1.9\, \mathrm{GeV}^{2}/\mathrm{fm}$ at $t=50\,\mathrm{fm}/\mathrm{c}$. The combined $\langle \hat{q} \rangle$ is dominated by the radiative $gg \rightarrow ggg$ contribution and starts at $\langle \hat{q} \rangle \approx 23\, \mathrm{GeV}^{2}/\mathrm{fm}$, falling to $\langle \hat{q} \rangle \approx 12.5\, \mathrm{GeV}^{2}/\mathrm{fm}$ at $t=3.5\,\mathrm{fm}/\mathrm{c}$. This indicates that the negligence of transverse momentum pick--up in radiative processes might indeed be an oversimplification.

The numbers for $\hat{q}$ found in this work are well within the range of values found by other theoretical energy loss schemes, though the comparison is difficult since $\hat{q}$ in these calculations often is a free parameter or related to free parameters. Fitting to experimental data the authors of \cite{Bass:2008rv} have found $\hat{q}_{0}$ for the central region of Au+Au at $\tau_{0} = 0.6\,\mathrm{fm/c}$, where conditions should be roughly comparable to the setup used in this section, to be ranging from $2.3\, \mathrm{GeV}^{2}/\mathrm{fm}$ based on the Higher Twist approach \cite{Majumder:2007ae}, over $4.1\, \mathrm{GeV}^{2}/\mathrm{fm}$ based on the approach by Arnold-Moore-Yaffe (AMY) \cite{Qin:2007zz}, up to $18.5\, \mathrm{GeV}^{2}/\mathrm{fm}$ based on the approach by Armesto-Salgado-Wiedemann (ASW) \cite{Renk:2006sx}. In \cite{Chen:2010te} the application of the Higher Twist approach to jet quenching data yields $\hat{q}_{0} \approx 3.2\, \mathrm{GeV}^{2}/\mathrm{fm}$ for a medium evolution based on BAMPS (employing $\tau_{0} = 0.3\,\mathrm{fm/c}$), while a hydro based medium evolution yields $\hat{q}_{0} \approx 0.9\, \mathrm{GeV}^{2}/\mathrm{fm}$ ($\tau_{0} = 0.6\,\mathrm{fm/c}$).

\section{Non--central Au+Au collisions at RHIC energy} \label{sec:non_central_RHIC}

In a previous work \cite{Fochler:2008ts} BAMPS has been applied to simulate elliptic flow and jet quenching at RHIC energies, for the first time using a consistent and fully pQCD--based microscopic parton transport model to approach both key observables of heavy ion collisions at RHIC within a common setup. As established in \cite{Xu:2007jv, Xu:2008av} the medium simulated in the parton cascade BAMPS exhibits a sizable degree of elliptic flow in agreement with experimental findings at RHIC, while showing a small ratio of shear viscosity to entropy $\eta / s$ \cite{Xu:2007ns}. The suppression of high--$p_{T}$ gluon jets in central, $b=0\,\mathrm{fm}$, collisions has been found to be roughly constant at $R_{AA}^{\mathrm{gluons}} \approx 0.053$ with radiative events $gg \rightarrow ggg$ being the dominant contribution to jet energy loss. This nuclear modification factor is in reasonable agreement with recent analytic results for the gluonic contribution to the nuclear $R_{AA}$ \cite{Wicks:2005gt}, though the suppression of gluon jets in BAMPS appears to be slightly stronger. However, we expect improved agreement in future studies when employing a carefully averaged $\langle b \rangle$ that will be better suited for comparisons to experimental data, having a centrality range even for the most central selections, than the strict $b=0\,\mathrm{fm}$ case.

In order to further test the energy loss as modeled by BAMPS and to allow for more extensive comparisons with experimental data and analytic models, our investigations need to be extended. One way to gain further insight is to study the evolution of jets in the simulations of Au+Au collisions including light quark degrees of freedom and to subsequently employ a fragmentation scheme that would allow for direct comparison with hadronic data. The consistent implementation of inelastic 2- and 3--particle interactions for all possible combinations of light quarks, light anti--quarks and gluons is challenging and we leave this to an upcoming work. Another way is to look at more differential observables, for instance to study the nuclear modification factor as a function of centrality or to study the dependence on the angle with respect to the reaction plane \cite{Renk:2008xq, Li:2009ti}.

Here we will present first results in this direction by discussing the nuclear modification factor as simulated by BAMPS for non--central Au + Au collisions at the RHIC energy of $\sqrt{s} = 200 \mathrm{AGeV}$ with a fixed impact parameter $b=7\,\mathrm{fm}$. This roughly corresponds to $20 \%$ to $30 \%$ experimental centrality. 

\begin{figure}[htbp]
  \begin{center}
    \includegraphics[width=14cm]{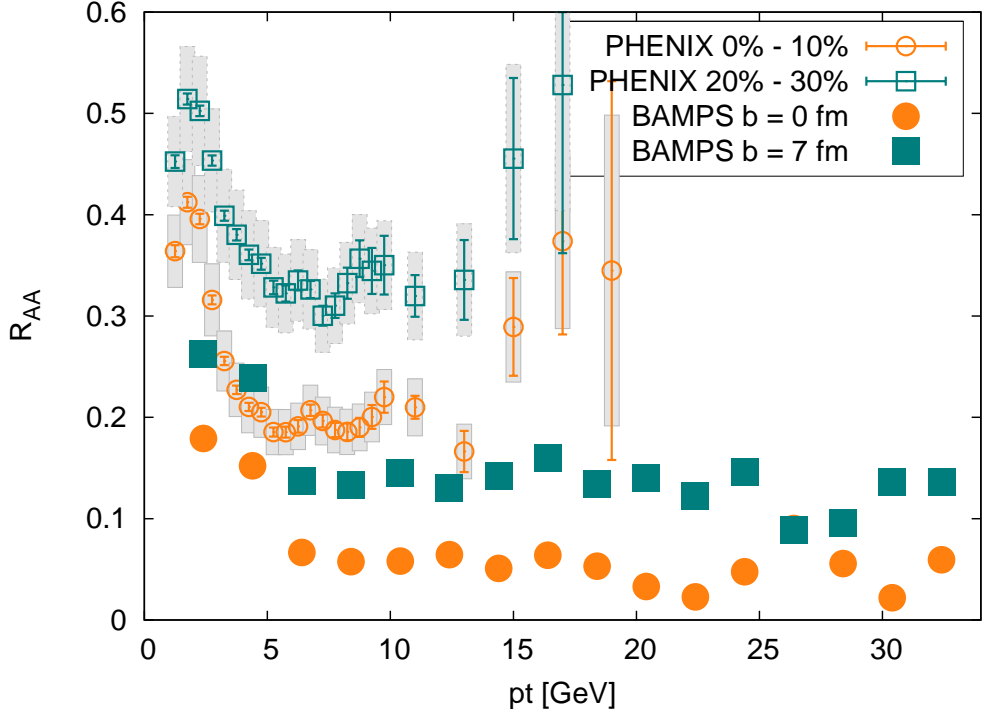}
    \caption{(Color online) Gluonic nuclear modification factor $R_{AA}$ at mid-rapidity ($y \,\epsilon\, [-0.5,0.5]$) as extracted from simulations of Au+Au collisions at 200~AGeV with fixed impact parameters $b=0\,\mathrm{fm}$ and $b=7\,\mathrm{fm}$. For comparison experimental results from PHENIX \cite{Adare:2008qa} for $\pi^{0}$ are shown for central ($0 \%$ - $10 \%$) and off--central ($20 \%$ - $30 \%$) collisions.}
    \label{fig:RAA_b7}
  \end{center}
\vspace{-0.5cm}
\end{figure}

For these simulations the initial gluon distributions are sampled according to a mini--jet model with a lower momentum cut-off $p_{0} = 1.4\,\mathrm{GeV}$ and a $K$--factor of $2$. The underlying nucleon-nucleon collisions follow a Glauber-model with a Wood-Saxon density profile and the results by Gl\"uck, Reya and Vogt \cite{Gluck:1994uf} are used as parton distribution functions. Quarks are discarded after sampling the initial parton distribution since currently a purely gluonic medium is considered. To model the freeze out of the simulated fireball, free streaming is applied to regions where the local energy density has dropped below a critical energy density $\varepsilon_{c}$. This setup has been successfully checked against experimental findings such as the distribution of transverse energy in rapidity and the flow parameter $v_{2}$ at various centralities in \cite{Xu:2007aa, Xu:2007jv, Xu:2008av}.  

The test particle method \cite{Xu:2004mz} is employed to ensure sufficient statistics and to allow for the resolution of adequate spatial length scales. For the calculations of the $b=7\,\mathrm{fm}$ events presented here, we use $N_{\text{test}}=220$. The size of the computational grid in the transverse direction is $\Delta x = \Delta y = 0.2\,\mathrm{fm}$ and in the longitudinal direction an adaptive grid linked to the number of particles in the cells is used that roughly corresponds to $\Delta \eta \approx 0.1 \div 0.2$.

Because of the steeply falling momentum spectrum of the initial gluon distribution the computational expense of calculating observables at high--$p_{T}$ becomes very high, rendering a brute force approach that consists in simulating a huge number of random events completely infeasible. Therefore, in an approach similar to the concept of importance sampling, only selected events are chosen for computation and a suitable weighting and reconstruction scheme is applied. A large number of initial spectra is characterized according to $X = \max \left( \left(p_{T}\right)_{i}\right)$, the maximum $p_{T}$ of the particles in an event in a given rapidity range. Subsequently $X$ is divided into bins of appropriate size and from each bin $j$ in $X$ a number $N_{j}$ of events is selected for simulation. The results are averaged over the $N_{j}$ events within a bin and finally combined with the appropriate weights $P_{j}$, where $P_{j}$ is the probability that in a simulated initial state there are one or more particles with their transverse momentum in the interval $\left[ \left(p_{T} \right)_{j-1}, \left(p_{T}\right)_{j} \right]$ but none with $p_{T} > \left(p_{T}\right)_{j}$. We use a bin size of $\Delta p_{T} = 1\,\mathrm{GeV}$ and for this work select up to $N_{j} = 40$ events per bin.

We have successfully tested this method by reconstructing simple analytic distributions such as $p(s)=e^{-s}$ and more importantly by reconstructing the initial mini--jet $p_{T}$--distribution from a small set of selected events.

Since within the mini--jet model employed in these simulations the scaled nucleon--nucleon $p_{T}$--spectrum is directly accessible, the usual definition of the nuclear modification factor
\begin{equation} \label{eq:RAA}
R_{AA}=\frac{d^{2}N_{AA}/dp_{T}dy}{T_{AA}d^{2}\sigma_{NN}/dp_{T}dy}
\end{equation}
simplifies to just the ratio of the final $p_{T}$--spectrum over the initial $p_{T}$--spectrum.

Figure \ref{fig:RAA_b7} shows the gluonic contribution to the nuclear modification factor $R_{AA}$ as extracted from simulations with BAMPS for fixed impact parameters $b=0\,\mathrm{fm}$ and $b=7\,\mathrm{fm}$. The critical freeze out energy density is $\varepsilon_{c} = 1.0\, \mathrm{GeV}/\mathrm{fm}^3$, see section \ref{sec:sensitivity} for results employing $\varepsilon_{c} = 0.6\, \mathrm{GeV}/\mathrm{fm}^3$. Most notably the quenching pattern for $b=7\,\mathrm{fm}$ is flat as for the central collisions but at $R_{AA}\approx 0.13$. This is qualitatively consistent with experimental observations since the measured shape of the nuclear modification factor for $\pi^{0}$ does not change significantly going from the centrality class $0 \%$ - $10 \%$ to $20 \%$ - $30 \%$ \cite{Adare:2008qa}.

A comparison in terms of the magnitude of the jet suppression for $b=7\,\mathrm{fm}$ is difficult since to the best of our knowledge there are no published results from analytic models that explicitly disentangle the gluon and quark contributions to non--central $R_{AA}$. Taking the ratio of non--central to central $R_{AA}$ presented in Fig. \ref{fig:RAA_b7} as a rough guess indicates that the decrease in quenching is more pronounced in BAMPS than in the experimental data. The ratio of the nuclear modification factor between central ($0 \%$ - $10 \%$) and more peripheral ($20 \%$ - $30 \%$) collisions is $\left. R_{AA}\right|_{0 \% - 10 \%} / \left. R_{AA}\right|_{20 \% - 30 \%} \approx 0.6$ for the experimental data, while for the BAMPS results $\left. R_{AA}\right|_{b=0\,\mathrm{fm}} / \left. R_{AA}\right|_{b=7\,\mathrm{fm}} \approx 0.4$. However, for thorough quantitative comparison one needs to address the question how fragmentation influences the relative gluon and quark contribution to the nuclear modification factor at different centralities. The issue of detailed quantitative comparison therefore needs to be settled once light quarks and a fragmentation scheme are included into the simulations.

\begin{figure}[htbp]
  \begin{center}
    \includegraphics[width=14cm]{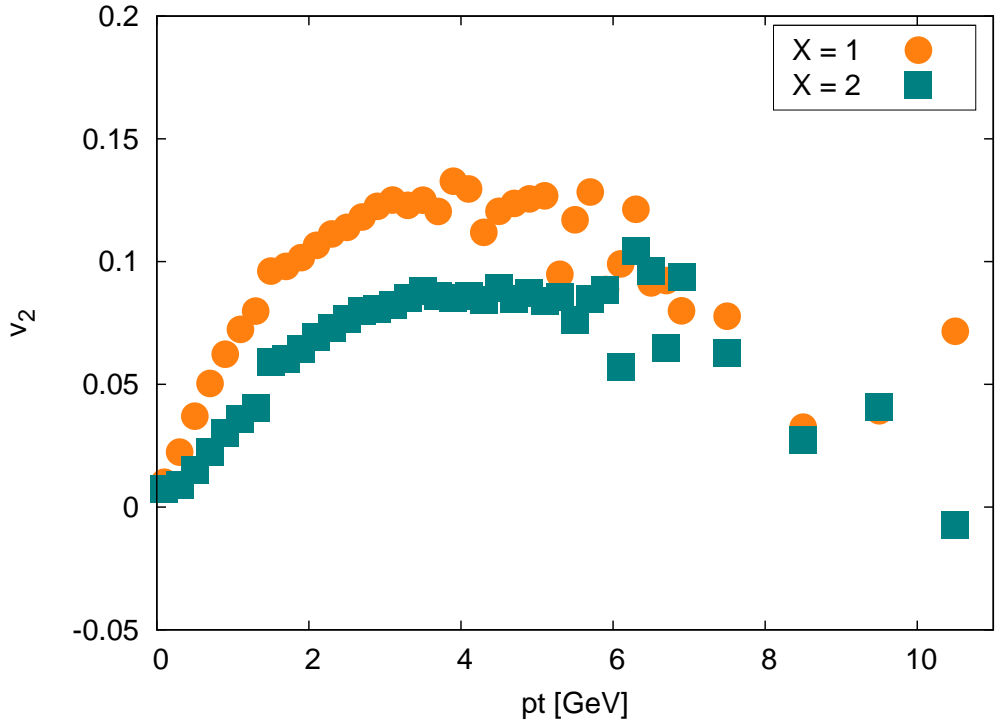}
    \caption{(Color online) Elliptic flow parameter $v_{2}$ for gluons as a function of transverse momentum $p_{T}$. Extracted from BAMPS calculations of a non--central $b=7\,\mathrm{fm}$ Au + Au collisions at 200~AGeV using $\varepsilon_{c}=0.6\, \mathrm{GeV}/\mathrm{fm}^3$. Shown are two cases: The regular scenario where $X=1$ and the scenario in which the LPM cut--off (\ref{eq:modified_theta_function}) is modified by a factor $X=2$.}
    \label{fig:v2_b7}
  \end{center}
\vspace{-0.5cm}
\end{figure}

To complement the investigations of $R_{AA}$ at a non--zero impact parameter, in Fig. \ref{fig:v2_b7} we present the elliptic flow parameter $v_{2}$ for gluons computed within BAMPS  at the same impact parameter $b=7\,\mathrm{fm}$ as used above. The elliptic flow in this figure is obtained from simulations using a critical energy density $\varepsilon_{c} = 0.6\, \mathrm{GeV}/\mathrm{fm}^3$ in order to be comparable to previous results. Going from $\varepsilon = 1\, \mathrm{GeV}/\mathrm{fm}^3$ to $\varepsilon = 0.6\, \mathrm{GeV}/\mathrm{fm}^3$ does not change the magnitude of the nuclear suppression factor $R_{AA}$ but slightly distorts its flatness, see the results marked ``$X=1$'' in Fig. \ref{fig:RAA_cutoffs}. Compared to previous studies \cite{Xu:2007jv, Xu:2008av} the elliptic flow results presented here extend to high--$p_{T}$ gluons, up to transverse momenta of roughly $p_{T} \approx 10\,\mathrm{GeV}$. Bearing in mind the limited statistics of this calculation, the $v_{2}$ of high--$p_{T}$ gluons is rising up to $p_{T} \approx 4\,\mathrm{GeV}$. Afterward, from about $p_{T} \approx 5\,\mathrm{GeV}$ on, the results indicate a decrease in the elliptic flow for high--$p_{T}$ gluons. This behavior is in good qualitative agreement with recent RHIC data \cite{Abelev:2008ed} that for charged hadrons shows $v_{2}$ to be rising up to $v_{2} \approx 0.15$ at $p_{T} \approx 3\,\mathrm{GeV}$ followed by a slight decrease.

\section{Sensitivity of the results on the LPM cut--off} \label{sec:sensitivity}

Every effective model comes with a set of free parameters whose choice can be motivated by physical arguments or by fits to experimental data. The most notable free parameter in the transport model BAMPS is the coupling strength $\alpha_{s}$ that has been fixed to the canonical value of $\alpha_{s} = 0.3$ throughout this work. The consequences of different choices of $\alpha_{s}$, especially on elliptic flow observables, have been studied in previous works \cite{Xu:2007jv,Xu:2008av}. In simulations of heavy ion collisions the freeze out energy density $\varepsilon_{c}$, see section \ref{sec:non_central_RHIC}, is also a parameter that can be adjusted within certain limits.

\begin{figure}[htbp]
  \begin{center}
    \includegraphics[width=14cm]{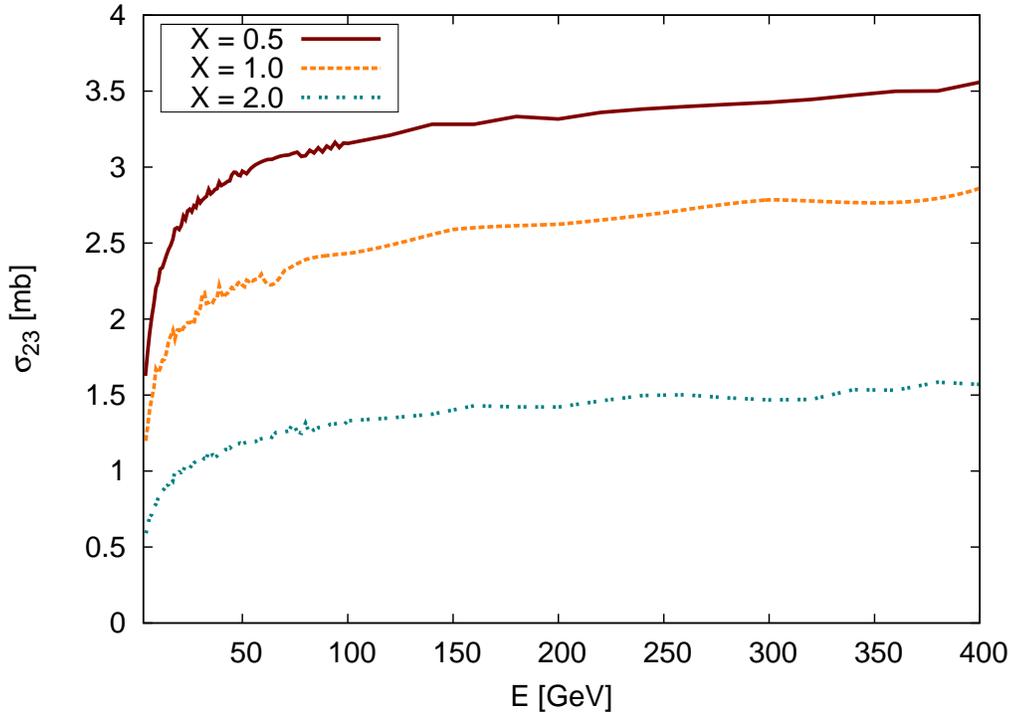}
    \caption{(Color online) Total cross section for $gg \rightarrow ggg$ processes involving a gluon jet with energy $E$ that traverses a thermal medium ($T=400\,\mathrm{MeV}$). Shown are three different values for the factor $X$ that modifies the effective LPM cut--off (\ref{eq:modified_theta_function}).}
    \label{fig:sigma_cutoffs}
  \end{center}
\vspace{-0.5cm}
\end{figure}

When investigating inelastic $gg \leftrightarrow ggg$ interactions within BAMPS there basically enters another parameter due to the effective modeling of the LPM effect via a cut-off. In the $gg \rightarrow ggg$ matrix element (\ref{eq:gg_to_ggg}) the cut-off is realized via a Theta function that essentially compares the formation time $\tau$ of the radiated gluon to the mean free path of the the parent gluon $\Lambda_{g}$ as discussed in section \ref{sec:BAMPS} in more detail. However, the argument underlying the distinction between coherent and incoherent processes via a threshold $\tau = \Lambda_{g}$ is of course a qualitative one. When effectively modeling the LPM effect via a cut--off the Theta function (\ref{eq:theta_function}) could therefore be replaced by a more general form
\begin{equation}
\label{eq:modified_theta_function}
\Theta\left( k_{\perp} - \frac{\gamma}{\Lambda_g} \right) \rightarrow \Theta\left( k_{\perp} - X \frac{\gamma}{\Lambda_g} \right)
\text{,}
\end{equation}
where $X$ is a real number not too far from 1.

In this section we explore the consequences on our results when modifying the LPM--cutoff (\ref{eq:theta_function}) by a factor $X$. Specifically we choose $X = 0.5$, $X=1$ (the usual choice) and $X=2$. This should provide a grasp on how sensitive the results for partonic energy loss and collective flow within BAMPS simulations are on the specific prescription for including the LPM effect.

The computation of total cross section for radiative $gg \rightarrow ggg$ processes involves an integral of the matrix element (\ref{eq:gg_to_ggg}) over the transverse momentum $k_{\perp}$. It is therefore straightforward that a larger $X$ in the cut--off (\ref{eq:modified_theta_function}) corresponds to a smaller total cross section. Though the actual dependence is not linear and indeed non--trivial, Fig. \ref{fig:sigma_cutoffs}, that shows the total cross section for $gg \rightarrow ggg$ processes in a gluonic medium with $T = 400\,\mathrm{MeV}$ for different choices of $X$, confirms the simple qualitative considerations. The change in the cross section naturally corresponds to a change in the rate for this process via $R_{23} = \langle n \sigma_{23} \rangle$ and thus in the mean free path between radiative processes $\Lambda_{23} = 1/R_{23}$.

\begin{figure}[htbp]
  \begin{center}
    \includegraphics[width=14cm]{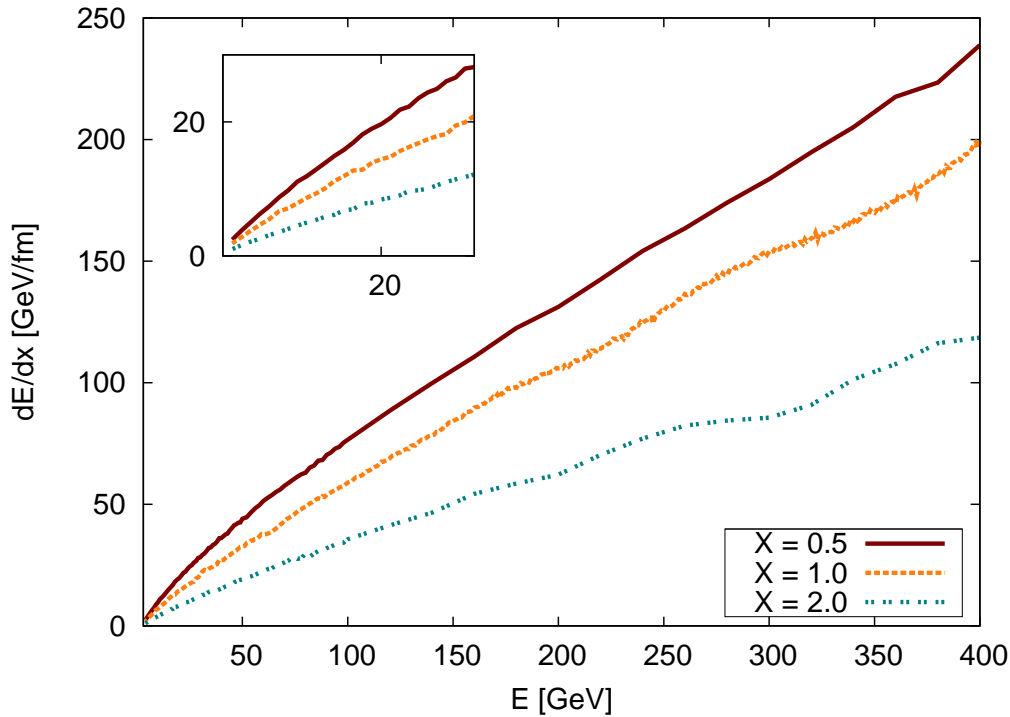}
    \caption{(Color online) Differential energy loss of a gluon jet with energy $E$ that traverses a thermal medium ($T=400\,\mathrm{MeV}$). Shown are three different values for the factor $X$ that modifies the effective LPM cut--off (\ref{eq:modified_theta_function}).}
    \label{fig:dEdx_all_cutoffs}
  \end{center}
\vspace{-0.5cm}
\end{figure}

\begin{figure}[htbp]
  \begin{center}
    \includegraphics[width=14cm]{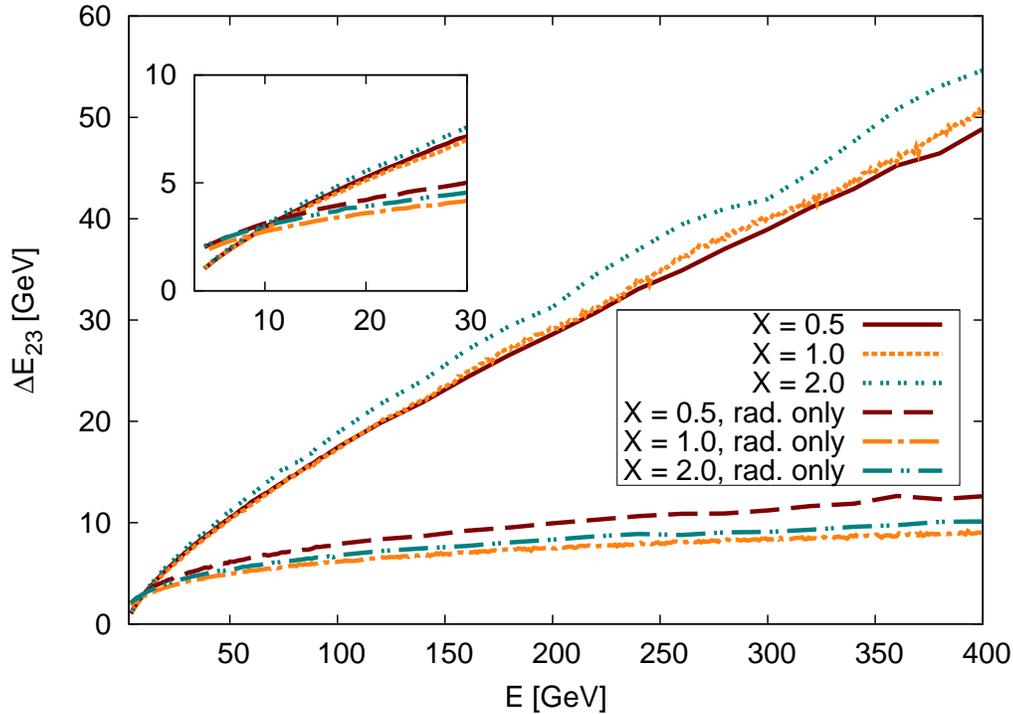}
    \caption{(Color online) Energy loss $\Delta E_{23}$ of a gluon jet in a single $gg \rightarrow ggg$ process ($T=400\,\mathrm{MeV}$). Two different cases are shown. One where $\Delta E = E^{\text{in}} - \max \left( E_{1}^{\text{out}}, E_{2}^{\text{out}}, E_{3}^{\text{out}} \right)$ (the upper group of lines) and one (labeled ``rad. only'', the lower group of lines) where $\Delta E = \omega$ is the energy of the radiated gluon. For each case three values for the parameter $X$ in the LPM cut--off (\ref{eq:modified_theta_function}) are explored.}
    \label{fig:deltaE_23_cutoffs}
  \end{center}
\vspace{-0.5cm}
\end{figure}

Correspondingly the differential energy loss $dE/dx$ is affected by a change in the parameter $X$ as shown in Fig. \ref{fig:dEdx_all_cutoffs}. Larger $X$ leads to a larger mean free path and thus a smaller energy loss per path length. That the change in $dE/dx$ is indeed mainly due to the change in the total cross section can be seen when comparing with Fig. \ref{fig:deltaE_23_cutoffs}, where it is shown that the effect of changes in $X$ on the energy lost in a single $gg \rightarrow ggg$ interaction is rather small. Also note that the ordering of the energy loss in a single interaction is not as straightforward as in the total cross section. This is due to the highly non--trivial impact of the LPM cut--off on the sampling of outgoing particle momenta in $gg \rightarrow ggg$ processes.

\begin{figure}[htbp]
  \begin{center}
    \includegraphics[width=14cm]{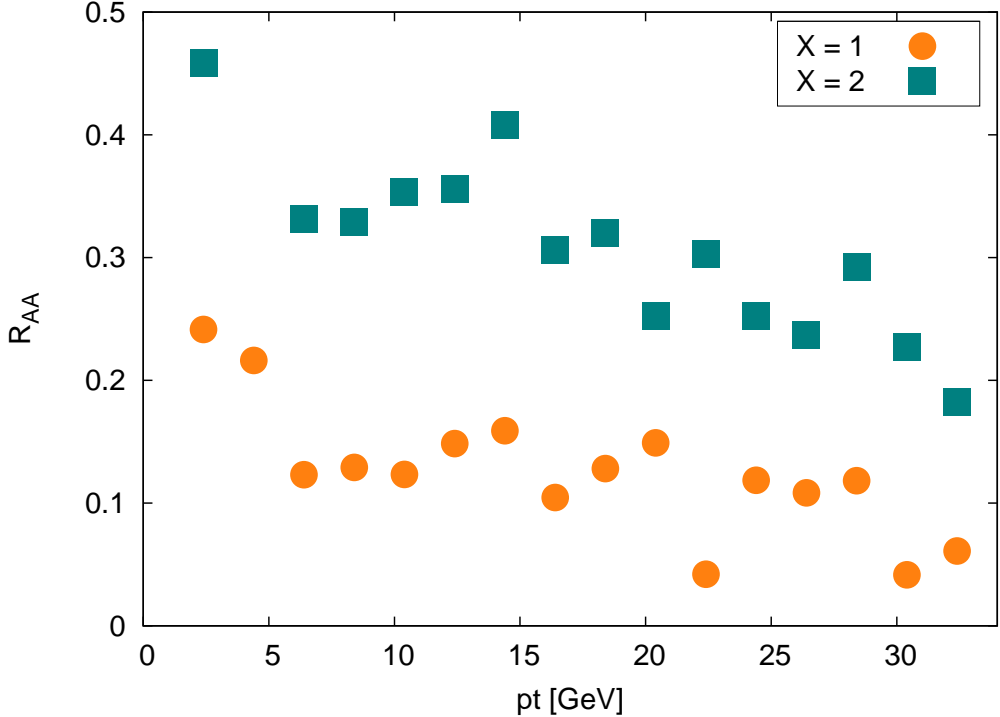}
    \caption{(Color online) Comparison of gluonic $R_{AA}$ for different values of the parameter $X$ in the effective LPM cut--off (\ref{eq:modified_theta_function}). $R_{AA}$ extracted at mid-rapidity ($y \,\epsilon\, [-0.5,0.5]$) for Au+Au collisions at 200~AGeV with fixed impact parameter $b=7\,\mathrm{fm}$. Critical freeze out energy density is $\varepsilon_{c} = 0.6\, \mathrm{GeV}/\mathrm{fm}^3$.}
    \label{fig:RAA_cutoffs}
  \end{center}
\vspace{-0.5cm}
\end{figure}

Finally we investigate the effect of a modified LPM cut-off (\ref{eq:modified_theta_function}) on observables in full simulations of heavy ion collisions as discussed in section \ref{sec:non_central_RHIC}. Due to limited computing resources we restrict ourselves to a comparison of the cases $X = 1$ and $X = 2$. As was to be expected from the change in the energy loss in a static medium (Fig. \ref{fig:dEdx_all_cutoffs}) comparing different cut-off parameters, the level of jet quenching is considerably reduced when going from $X = 1$ to $X = 2$, see Fig. \ref{fig:RAA_cutoffs}. Note that in the calculations for Fig. \ref{fig:RAA_cutoffs} we have used a freeze out energy density of $\varepsilon_{c} = 0.6\, \mathrm{GeV}/\mathrm{fm}^3$, which leads to a slight linear decrease of $R_{AA}$ towards high $p_{T}$ as compared to the results with $\varepsilon_{c} = 1.0\, \mathrm{GeV}/\mathrm{fm}^3$ in Fig. \ref{fig:RAA_b7}. The elliptic flow as seen in Fig. \ref{fig:v2_b7} is reduced by roughly 30\% to 35\% when going to $X=2$, but the qualitative features as a function of $p_{T}$ remain unaffected.

\section{Summary} \label{sec:summary}

In this work we have presented detailed results on the energy loss mechanisms for high--energy gluons within the microscopic parton transport model BAMPS. To this end we have investigated the evolution of high energy gluons within thermal and static media of gluons. This setup is known as the ``brick problem'' and is proposed by the TECHQM collaboration as means of comparing results from different parton cascade models and Monte Carlo calculations. We have discussed collisional energy loss as well as the contribution of radiative processes implemented in BAMPS via the Gunion--Bertsch matrix element (\ref{eq:gg_to_ggg}). The inelastic $gg \rightarrow ggg$ processes are found to be the dominant source of energy loss for high energy gluons in computations within the BAMPS model resulting in a strong differential energy loss that rises almost linearly with the jet energy. The strong mean energy loss in $gg \rightarrow ggg$ processes is due to a heavy tail in the $\Delta E$ distribution for single interactions, caused by the phase space configurations of outgoing particles dictated by the Gunion-Bertsch matrix element (\ref{eq:gg_to_ggg}) in combination with the effective LPM cutoff (\ref{eq:theta_function}). This prefers the emission of radiated gluons into the backward hemisphere with energies that in the center of momentum frame are comparable to that of the remaining outgoing particles, while they are small in the laboratory frame. The jet energy in these cases can be split into two particles yielding a large energy loss.

For a purely gluonic medium with $T=400\,\mathrm{MeV}$ the transport parameter $\hat{q}$ as defined in (\ref{eq:qhat_def}) and (\ref{eq:qhat_from_integral}) stemming from binary $gg \rightarrow gg$ interactions is found to be roughly constant at $\hat{q} = 2.3\, \mathrm{GeV}^{2}/\mathrm{fm}$. When including inelastic $gg \leftrightarrow ggg$ processes, $\hat{q}$ as a measure of the accumulated transverse momentum exhibits a stronger dependence on the path length and is much larger than for elastic interactions, $\hat{q} = 12 \div 23\, \mathrm{GeV}^{2}/\mathrm{fm}$.

Extending our investigations of jet quenching in central Au + Au collisions at RHIC energies \cite{Fochler:2008ts}, we have presented the gluonic contribution to the nuclear modification factor for non--central $b=7\,\mathrm{fm}$ collisions simulated in BAMPS. The gluonic $R_{AA}$ in these off--central events is found to be flat at $R_{AA} \approx 0.13$ over a wide range in $p_{T}$ for a critical energy density of $\varepsilon_{c} = 1.0\, \mathrm{GeV}/\mathrm{fm}^3$, being in qualitative agreement with the experimental results. Results employing $\varepsilon_{c} = 0.6\, \mathrm{GeV}/\mathrm{fm}^3$ exhibit a slight linear decrease in $R_{AA}$ towards high transverse momenta.

Since our transport approach allows for the simultaneous investigation of high--$p_{T}$ observables and bulk properties of the medium, we have also studied the elliptic flow parameter $v_{2}$ for gluons up to roughly $10\,\mathrm{GeV}$. $v_{2}$ peaks at a $p_{T}\approx 4 \div 5\,\mathrm{GeV}$ and slowly drops towards larger transverse momenta.

We have explored the sensitivity of the results on our implementation of the LPM effect by varying the momentum space cut-off that enters all computations involving the matrix element $M_{gg \rightarrow ggg}$ for the inelastic processes included in BAMPS. Changing this cut--off mainly affects the total cross section for radiative $gg \rightarrow ggg$ processes, leading to changes in the energy loss per mean free path. The nuclear modification factor is found to be more sensitive on the specific implementation of this cut--of than the elliptic flow.

The characteristics of the strongly interacting, but still fully pQCD based, medium within the BAMPS description will be studied in further detail and in improved quantitative comparison to experimental data in upcoming works including light quark degrees of freedom. This of course needs to be accompanied by a hadronization scheme that allows for a more direct connection to hadronic observables. While a consistent modeling of low--$p_{T}$ hadronization needs careful consideration, the application of fragmentation functions to the high--$p_{T}$ sector will be straightforward once light quarks are included.

The application of the transport model BAMPS to heavy quark, i.e. charm and bottom, elliptic flow and quenching will provide further valuable insight into the underlying mechanisms and the importance of a careful treatment of the fireball dynamics. Such studies are underway and will complement the results presented in this work. Recently an implementation of radiative processes based on Gunion--Bertsch type matrix elements has been successfully applied to heavy quark $v_{2}$ and $R_{AA}$ \cite{Gossiaux:2010yx}, indicating promising perspectives for the implementation within the established BAMPS framework.

Furthermore, in future projects we will study the medium response to high--$p_{T}$ particles that deposit energy into the medium created in heavy ion collisions via energy loss mechanisms. This will for example allow us to investigate the possible formation of so called mach cones. First studies \cite{Bouras:2009nn} have already very successfully demonstrated that BAMPS offers the ability to describe collective shock phenomena in a viscous hydrodynamic medium.

\section*{Acknowledgments}

This work has been supported by the Helmholtz International Center for FAIR within the framework of the LOEWE program launched by the State of Hesse. The simulations have been performed at the Center for Scientific Computing (CSC) at the Goethe University Frankfurt.

\appendix
\section{Typical phase space configurations in $gg \rightarrow ggg$ processes}
\label{sec:app_typical_configurations}

The characteristics of the kinematics in $gg \rightarrow ggg$ processes dictated by the Gunion--Bertsch matrix element in combination with the LPM cut--off (\ref{eq:gg_to_ggg}) can be studied by identifying typical regions of the phase space and by selecting on these regions. Respecting energy and momentum conservation, the kinematics of the three outgoing particles can be described by $6$ independent parameters. One such possible choice would be the set ($E_{1}^{\prime}$, $E_{3}^{\prime}$, $\cos( \theta_{1})$, $\cos( \theta_{3})$, $\phi_{1}$, $\phi_{3}$), where $E_{1}^{\prime}$, $E_{3}^{\prime}$ are the energies, $\theta_{1}$ and $\theta_{3}$ are the angles with respect to the incoming momentum $\vec{p}_{\text{jet}}^{\,\prime}$ and $\phi_{1}$ and $\phi_{3}$ are the azimuthal angles of the outgoing particles $1$ and $3$ (the emitted gluon), with all values being taken in the CM frame. The corresponding values for the outgoing particle $2$ can then be inferred from momentum conservation.

Another choice would be to replace $\cos( \theta_{1})$ and $\cos( \theta_{3})$ by the momentum transfers $q_{\perp}$ and $k_{\perp}$ as directly given in (\ref{eq:gg_to_ggg}). Note however, that this choice hides the information whether $\cos( \theta_{1})$ and $\cos( \theta_{3})$ are larger or smaller than zero, i.e. whether particles $1$ and $3$ are emitted in the forward or in the backward direction. Finally, also replacing $E_{3}^{\prime}$ by $y$, the rapidity of the emitted gluon, would yield a set of parameters that are closest to matrix element (\ref{eq:gg_to_ggg}).

For the purpose of this discussion we use the set ($E_{1}^{\prime}$, $E_{3}^{\prime}$, $q_{\perp}$, $k_{\perp}$, $\phi_{1}$, $\phi_{3}$) and throw in additional information by looking at the signs of $\cos( \theta_{1})$ and $\cos( \theta_{3})$ as needed. We ignore any dependence on the azimuthal angles $\phi_{1}$, $\phi_{3}$ and note that for a fixed $q_{\perp}$ the matrix element (\ref{eq:gg_to_ggg}) gives a $k_{\perp}$ that is typically on the order of $q_{\perp}$. Thus we select events according to $E_{1}^{\prime}$, $E_{3}^{\prime}$ and $q_{\perp}$.

\begin{figure}[hpt]
  \begin{center}
    \includegraphics[width=14cm]{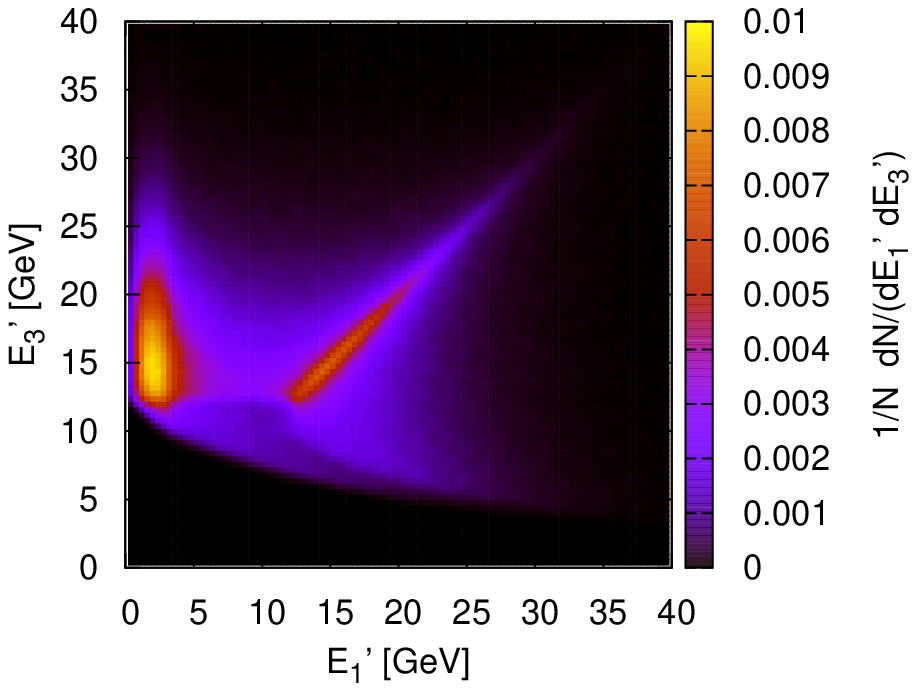}
    \includegraphics[width=14cm]{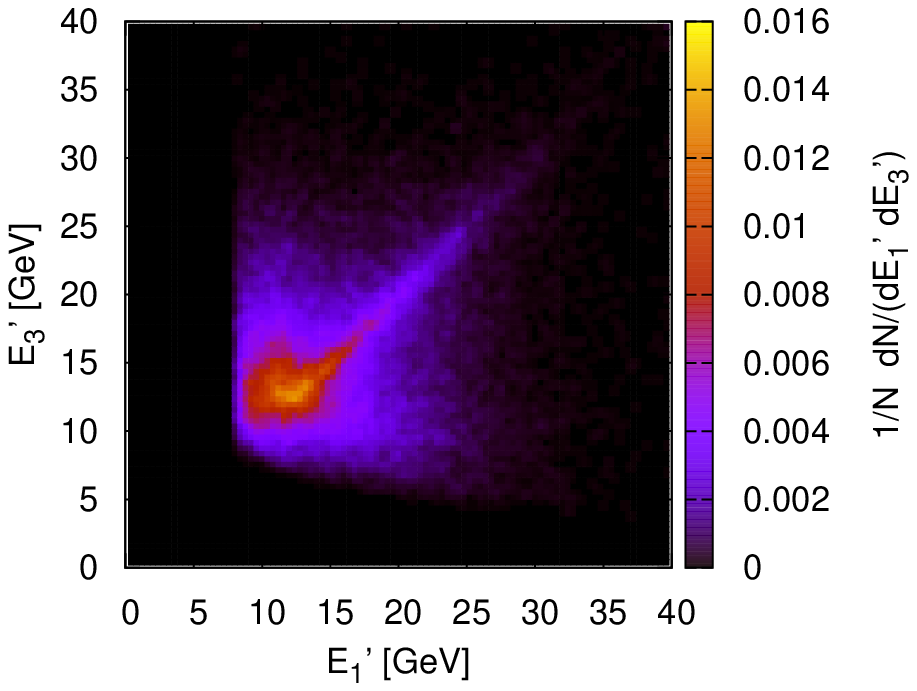}
    \caption{(Color online) Probability distribution of the outgoing energies $E_{1}^{\prime}$ and $E_{3}^{\prime}$ in the center of momentum frame (CM) for $gg \rightarrow ggg$ processes given a certain cut in the momentum transfer $q_{\perp}$, where $E_{3}^{\prime}$ is the energy of the emitted gluon. Jet energy $E=400\,\mathrm{GeV}$, medium temperature $T=400\,\mathrm{MeV}$.\newline
    Upper panel: $0\,\mathrm{GeV} \leq q_{\perp} \leq 3\,\mathrm{GeV}$. Lower panel: $8\,\mathrm{GeV} \leq q_{\perp} \leq 12\,\mathrm{GeV}$.}
    \label{fig:E3_vs_E1_qt_cut}
  \end{center}
\vspace{-0.5cm}
\end{figure}

Considering an $E = 400\,\mathrm{GeV}$ jet--like gluon inside a thermal medium with $T = 400\,\mathrm{MeV}$, we select events having a low $q_{\perp}$, $0\,\mathrm{GeV} \leq q_{\perp} \leq 3\,\mathrm{GeV}$, and events having a rather high $q_{\perp}$, $8\,\mathrm{GeV} \leq q_{\perp} \leq 12\,\mathrm{GeV}$. Figure \ref{fig:E3_vs_E1_qt_cut} then shows the color coded correlations between $E_{1}^{\prime}$ and $E_{3}^{\prime}$.

For low $q_{\perp}$, $0\,\mathrm{GeV} \leq q_{\perp} \leq 3\,\mathrm{GeV}$, two distinct regions in the $E_{1}^{\prime}$--$E_{3}^{\prime}$--plane are visible. The energy of the emitted gluon is quite high in all cases due to the strong preference of events with $y < 0$ caused by the LPM cut--off as discussed in section \ref{sec:BAMPS}. One region features small $E_{1}^{\prime}$, comparable to $q_{\perp}$ while the other regions features large $E_{1}^{\prime} \approx E_{3}^{\prime}$. For large $q_{\perp}$, $8\,\mathrm{GeV} \leq q_{\perp} \leq 12\,\mathrm{GeV}$ only one distinct region emerges with both $E_{1}^{\prime} \approx E_{3}^{\prime}$ large.

Table \ref{tab:selected_regions} lists the mean energy loss for events within these kinematical regions, differentiating between emission of particle $1$ into the forward direction ($\cos(\theta_{1}) > 0$) and into the backward direction ($\cos(\theta_{1}) < 0$). Additionally the relative abundance of events within these regions relative to all events is given.

\begin{table}
 \begin{tabular}{ | l | l | c | l | l| }
 \hline
 \multicolumn{5}{| c |}{$0 \leq q_{\perp} \leq 3$} \\
 \hline
 \hline
 \multirow{3}{*}{$0 \leq E_{1}^{\prime} \leq 5$}  & \multirow{3}{*}{$12 \leq E_{3}^{\prime} \leq 20$}  &  $\cos(\theta_{1}) < 0$  &  $7.5\,\%$  &  $\langle \Delta E \rangle \approx 13\,\mathrm{GeV}$ \\
   &   &  $\cos(\theta_{1}) > 0$  &  $7.5\,\%$  &  $\langle \Delta E \rangle \approx 61\,\mathrm{GeV}$ \\
   &   &  $-1 \leq \cos(\theta_{1}) \leq 1$  &  $15.0\,\%$  &  $\langle \Delta E \rangle \approx 37\,\mathrm{GeV}$ \\
 \hline
 \hline
  \multirow{3}{*}{$12 \leq E_{1}^{\prime} \leq 20$}  &  \multirow{3}{*}{$12 \leq E_{3}^{\prime} \leq 20$}  &  $\cos(\theta_{1}) < 0$  &  $0.4\,\%$  &  $\langle \Delta E \rangle \approx 7.7\,\mathrm{GeV}$ \\
  &  &  $\cos(\theta_{1}) > 0$  &  $13.6\,\%$  &  $\langle \Delta E \rangle \approx 30.7\,\mathrm{GeV}$ \\
  &  &  $-1 \leq \cos(\theta_{1}) \leq 1$  &  $14.0\,\%$  &  $\langle \Delta E \rangle \approx 30.0\,\mathrm{GeV}$ \\
 \hline   
 \multicolumn{5}{ c }{} \\
 \hline
 \multicolumn{5}{| c |}{$8 \leq q_{\perp} \leq 12$} \\
 \hline
 \hline
 \multirow{3}{*}{$8 \leq E_{1}^{\prime} \leq 15$}  & \multirow{3}{*}{$8 \leq E_{3}^{\prime} \leq 15$}  &  $\cos(\theta_{1}) < 0$  &  $0.34\,\%$  &  $\langle \Delta E \rangle \approx 105\,\mathrm{GeV}$ \\
   &   &  $\cos(\theta_{1}) > 0$  &  $0.56\,\%$  &  $\langle \Delta E \rangle \approx 135\,\mathrm{GeV}$ \\
   &   &  $-1 \leq \cos(\theta_{1}) \leq 1$  &  $0.9\,\%$  &  $\langle \Delta E \rangle \approx 123\,\mathrm{GeV}$ \\
 \hline
 \end{tabular} 
 \caption{Mean energy loss for given cuts in $q_{\perp}$, $E_{1}^{\prime}$ and $E_{3}^{\prime}$, cf. Fig. \ref{fig:E3_vs_E1_qt_cut}. The percentage given in column $4$ corresponds to the fraction of all events within these cuts relative to the total number of events. For reasons of readability the unit statement $GeV$ is omitted for $q_{\perp}$, $E_{1}^{\prime}$ and $E_{3}^{\prime}$.}
 \label{tab:selected_regions}
\end{table} 


\FloatBarrier

\section{Illustration of randomly selected $gg \rightarrow ggg$ events}
\label{sec:app_examples}

\begin{figure}[htb]
\centering
  \subfloat[$\Delta E = 0.67\,\mathrm{GeV}$,\newline
  $\cos(\Theta_{1}) = 0.998$, $p_{1}^{\,\prime} = (22.3, 22.2, 1.4, 0.0)$\newline
  $\cos(\Theta_{2}) = -0.487$, $p_{2}^{\,\prime} = (0.2, -0.1, 0.2, 0.0)$\newline
  $\cos(\Theta_{3}) = -0.997$, $p_{3}^{\,\prime} = (22.2, -22.1, -1.6, -0.0)$\newline
  $(E_{1}^{\text{lab}}, E_{2}^{\text{lab}}, E_{3}^{\text{lab}}) = (399.3, 1.0, 1.4)\,\mathrm{GeV}$\newline
  $q_{\perp} = 1.43\,\mathrm{GeV}$, $k_{\perp} = 1.60\,\mathrm{GeV}$]
  {
    \label{fig:illustrateVectors_CMS:a}
    \includegraphics[width=0.48\linewidth]{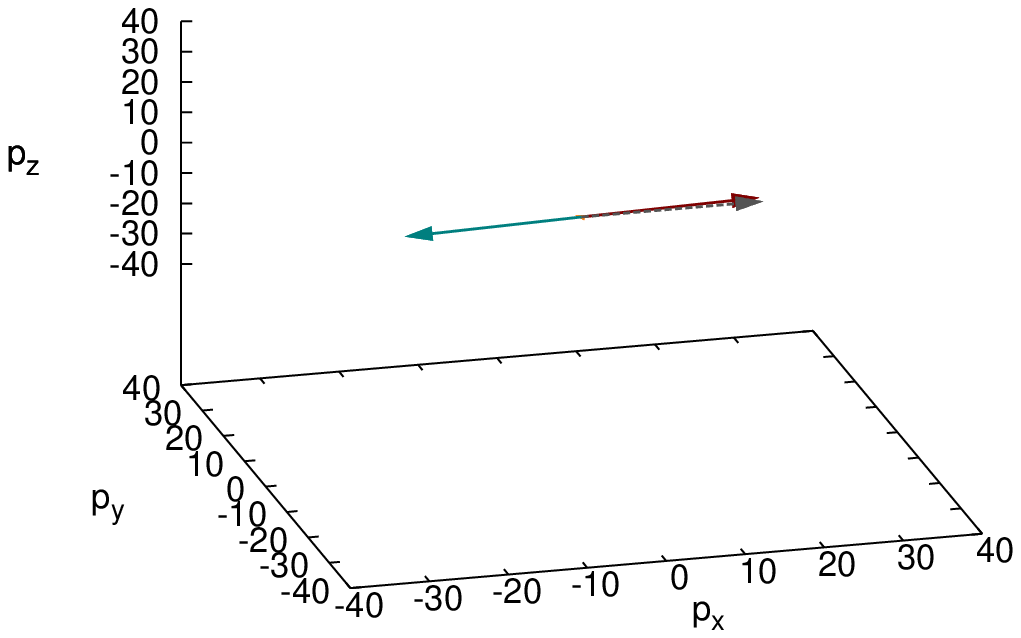}
  }
  \subfloat[$\Delta E = 206.43\,\mathrm{GeV}$,\newline
  $\cos(\Theta_{1}) = -0.134$, $p_{1}^{\,\prime} = (30.2, -4.0, 29.9, -0.0)$\newline
  $\cos(\Theta_{2}) = 0.829$, $p_{2}^{\,\prime} = (5.3, 4.4, 2.2, 2.0)$\newline
  $\cos(\Theta_{3}) = -0.011$, $p_{3}^{\,\prime} = (32.1, -0.3, -32.1, -2.0)$\newline
  $(E_{1}^{\text{lab}}, E_{2}^{\text{lab}}, E_{3}^{\text{lab}}) = (152.5, 56.9, 193.6)\,\mathrm{GeV}$\newline
  $q_{\perp} = 29.89\,\mathrm{GeV}$, $k_{\perp} = 32.13\,\mathrm{GeV}$]
  {
    \label{fig:illustrateVectors_CMS:b}
    \includegraphics[width=0.48\linewidth]{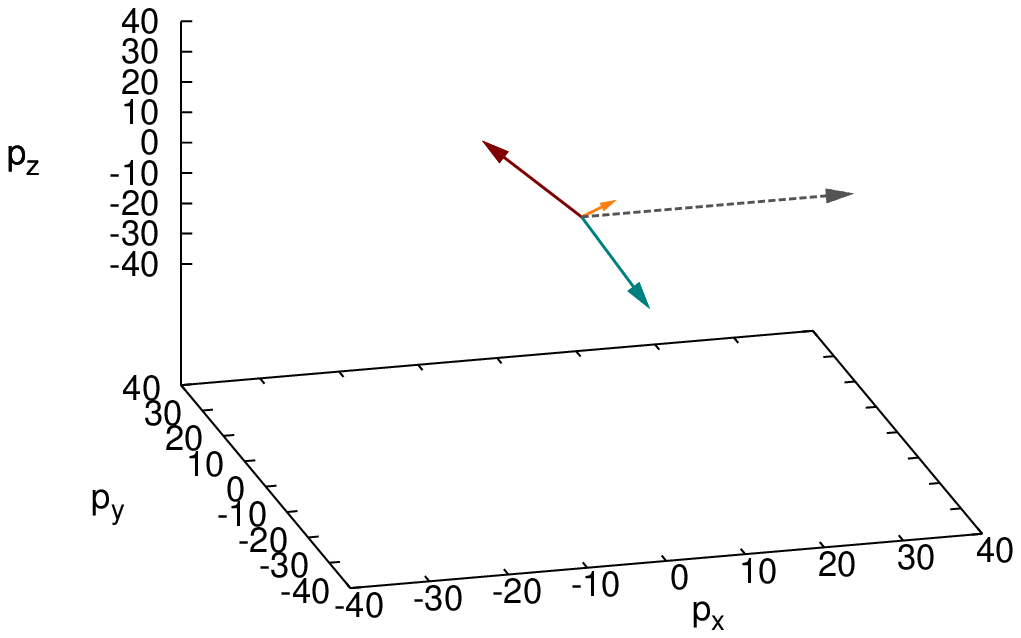}
  } \\[10pt]
  \subfloat[$\Delta E = 26.01\,\mathrm{GeV}$,\newline
  $\cos(\Theta_{1}) = -0.365$, $p_{1}^{\,\prime} = (4.3, -1.6, 4.0, 0.0)$\newline
  $\cos(\Theta_{2}) = 1.000$, $p_{2}^{\,\prime} = (22.8, 22.8, 0.3, 0.5)$\newline
  $\cos(\Theta_{3}) = -0.980$, $p_{3}^{\,\prime} = (21.6, -21.2, -4.3, -0.5)$\newline
  $(E_{1}^{\text{lab}}, E_{2}^{\text{lab}}, E_{3}^{\text{lab}}) = (21.8, 374.0, 6.0)\,\mathrm{GeV}$\newline
  $q_{\perp} = 4.02\,\mathrm{GeV}$, $k_{\perp} = 4.34\,\mathrm{GeV}$]
  {
    \label{fig:illustrateVectors_CMS:c}
    \includegraphics[width=0.48\linewidth]{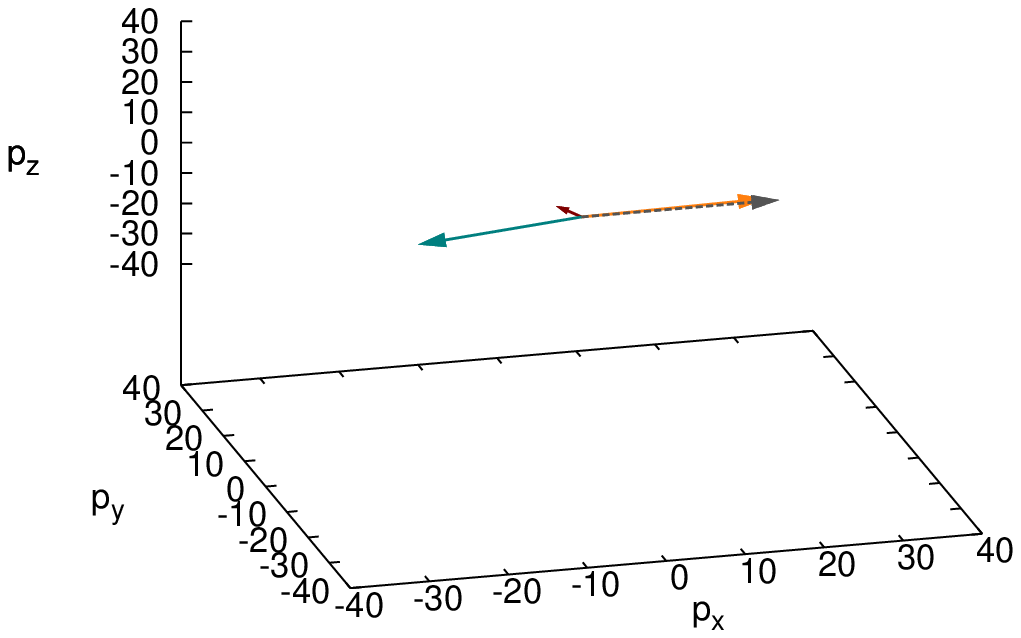}
  }
  \subfloat[$\Delta E = 117.01\,\mathrm{GeV}$,\newline
  $\cos(\Theta_{1}) = 0.998$, $p_{1}^{\,\prime} = (23.0, 23.0, 1.5, -0.0)$\newline
  $\cos(\Theta_{2}) = 0.976$, $p_{2}^{\,\prime} = (9.4, 9.1, 1.7, 1.1)$\newline
  $\cos(\Theta_{3}) = -0.994$, $p_{3}^{\,\prime} = (32.3, -32.1, -3.3, -1.1)$\newline
  $(E_{1}^{\text{lab}}, E_{2}^{\text{lab}}, E_{3}^{\text{lab}}) = (283.0, 112.7, 9.2)\,\mathrm{GeV}$\newline
  $q_{\perp} = 1.54\,\mathrm{GeV}$, $k_{\perp} = 3.45\,\mathrm{GeV}$]
  {
    \label{fig:illustrateVectors_CMS:d}
    \includegraphics[width=0.48\linewidth]{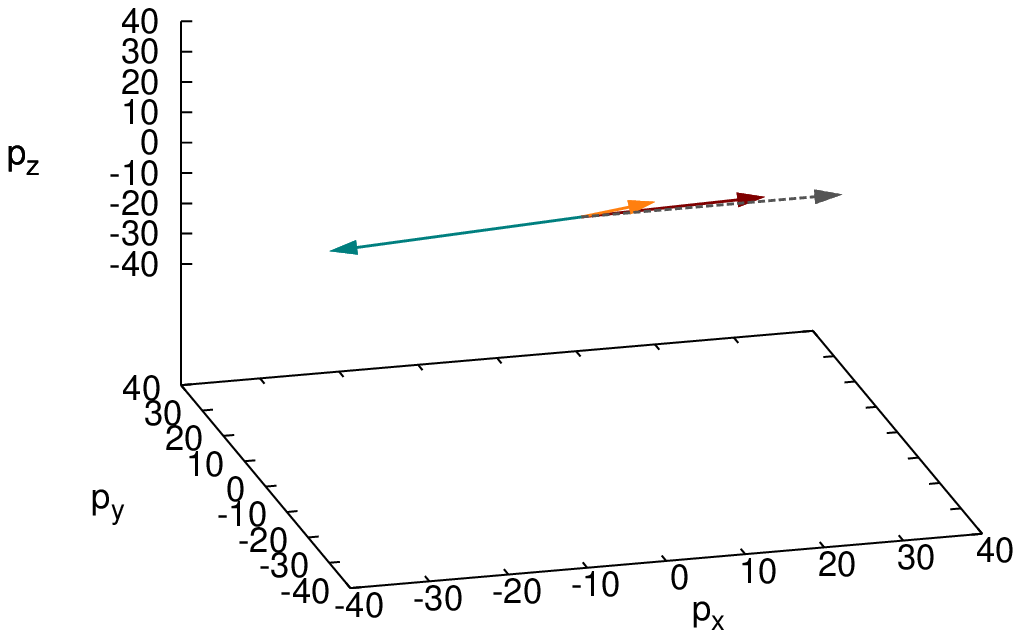}
  }

  \caption{(Color online) Randomly selected $gg \rightarrow ggg$ events involving a gluon jet with $E=400\,\mathrm{GeV}$ (lab system) displayed in the CM frame ($T=400\,\mathrm{MeV}$). All events are rotated such that the incoming jet momentum (CM) points along the positive x-direction and that the outgoing momentum $\vec{p}_1^{\,\prime}$ is in the x-y-plane. All kinematical values are given in the CM frame, execpt for the energies of the outgoing particles in the lab frame, $E_{1}^{\text{lab}}$, $E_{2}^{\text{lab}}$, $E_{3}^{\text{lab}}=\omega$. Part 1: Events 1-4 out of 8.
Dark red: $\vec{p}_{1}^{\,\prime}$, Orange: $\vec{p}_{2}^{\,\prime}$, Blue: $\vec{p}_{3}^{\,\prime}$ (radiated), Gray: $\vec{p}_{\text{jet}}^{\,\prime}$ (incoming).}
  \label{fig:illustrateVectors_CMS1}
\end{figure}

\begin{figure}
  \subfloat[$\Delta E = 54.26\,\mathrm{GeV}$,\newline
  $\cos(\Theta_{1}) = 0.861$, $p_{1}^{\,\prime} = (31.5, 27.1, 16.0, -0.0)$\newline
  $\cos(\Theta_{2}) = 0.983$, $p_{2}^{\,\prime} = (3.0, 3.0, 0.5, 0.2)$\newline
  $\cos(\Theta_{3}) = -0.876$, $p_{3}^{\,\prime} = (34.4, -30.1, -16.5, -0.2)$\newline
  $(E_{1}^{\text{lab}}, E_{2}^{\text{lab}}, E_{3}^{\text{lab}}) = (345.7, 35.2, 22.6)\,\mathrm{GeV}$\newline
  $q_{\perp} = 16.03\,\mathrm{GeV}$, $k_{\perp} = 16.54\,\mathrm{GeV}$]
  {
    \label{fig:illustrateVectors_CMS:e}
    \includegraphics[width=0.48\linewidth]{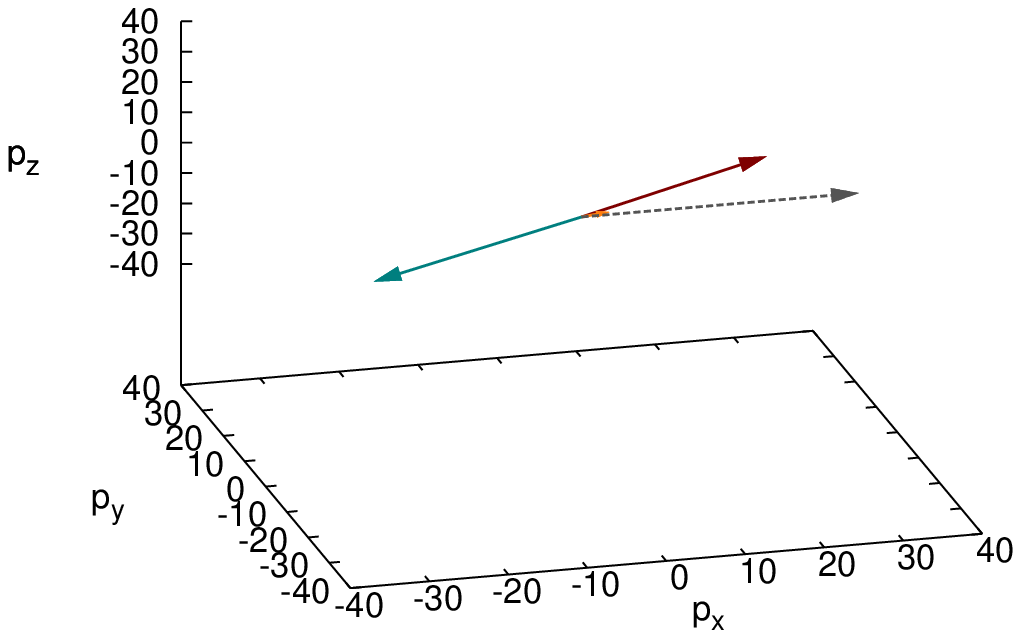}
  }
  \subfloat[$\Delta E = 71.64\,\mathrm{GeV}$,\newline
  $\cos(\Theta_{1}) = 0.997$, $p_{1}^{\,\prime} = (17.4, 17.4, 1.2, -0.0)$\newline
  $\cos(\Theta_{2}) = -0.277$, $p_{2}^{\,\prime} = (8.0, -2.2, -1.2, 7.6)$\newline
  $\cos(\Theta_{3}) = -0.894$, $p_{3}^{\,\prime} = (17.0, -15.2, -0.0, -7.6)$\newline
  $(E_{1}^{\text{lab}}, E_{2}^{\text{lab}}, E_{3}^{\text{lab}}) = (328.4, 50.9, 22.2)\,\mathrm{GeV}$\newline
  $q_{\perp} = 1.25\,\mathrm{GeV}$, $k_{\perp} = 7.60\,\mathrm{GeV}$]
  {
    \label{fig:illustrateVectors_CMS:f}
    \includegraphics[width=0.48\linewidth]{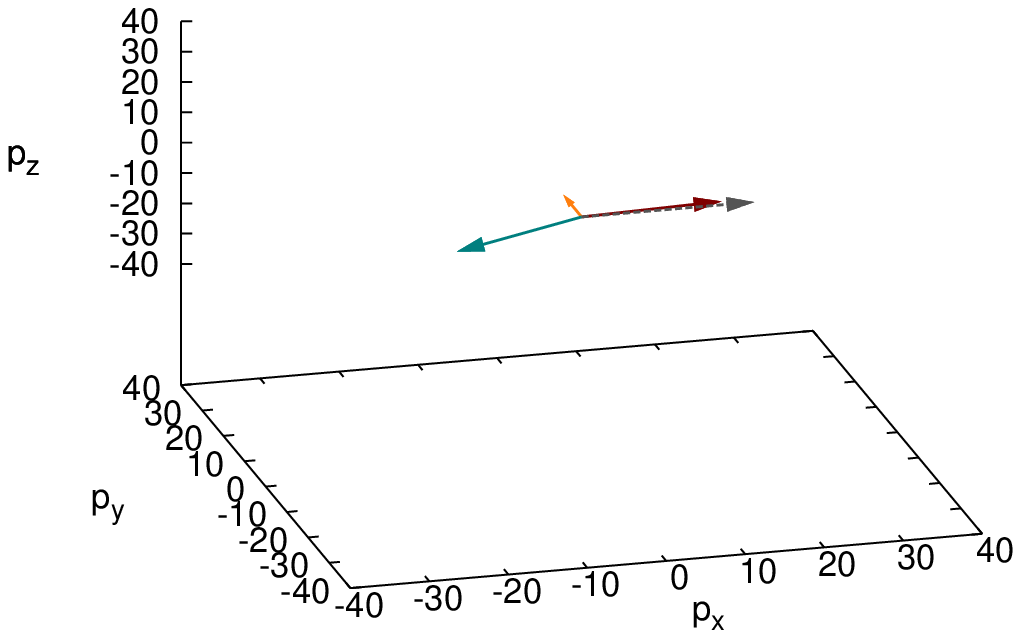}
  } \\[10pt]
  \subfloat[$\Delta E = 14.15\,\mathrm{GeV}$,\newline
  $\cos(\Theta_{1}) = -0.133$, $p_{1}^{\,\prime} = (1.4, -0.2, 1.3, 0.0)$\newline	
  $\cos(\Theta_{2}) = 0.999$, $p_{2}^{\,\prime} = (16.6, 16.6, -0.1, 0.7)$\newline
  $\cos(\Theta_{3}) = -0.996$, $p_{3}^{\,\prime} = (16.5, -16.4, -1.2, -0.7)$\newline
  $(E_{1}^{\text{lab}}, E_{2}^{\text{lab}}, E_{3}^{\text{lab}}) = (14.0, 385.8, 1.0)\,\mathrm{GeV}$\newline
  $q_{\perp} = 1.35\,\mathrm{GeV}$, $k_{\perp} = 1.39\,\mathrm{GeV}$]
  {
    \label{fig:illustrateVectors_CMS:g}
    \includegraphics[width=0.48\linewidth]{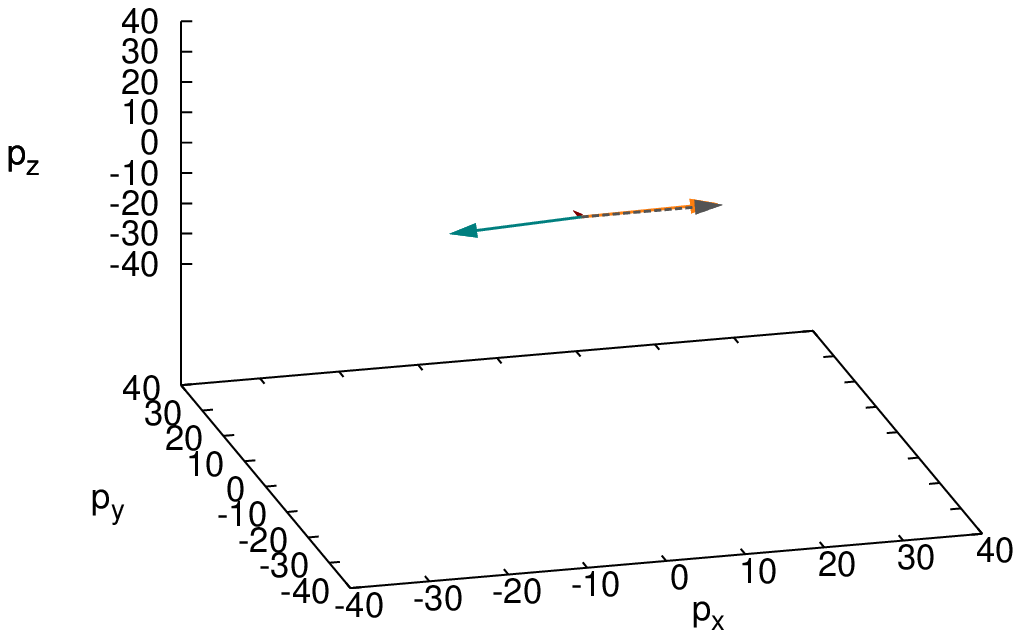}
  }
  \subfloat[$\Delta E = 41.35\,\mathrm{GeV}$,\newline
  $\cos(\Theta_{1}) = 0.953$, $p_{1}^{\,\prime} = (24.1, 23.0, 7.3, 0.0)$\newline
  $\cos(\Theta_{2}) = -0.881$, $p_{2}^{\,\prime} = (22.3, -19.6, -9.3, 4.9)$\newline
  $\cos(\Theta_{3}) = -0.535$, $p_{3}^{\,\prime} = (6.3, -3.4, 2.0, -4.9)$\newline
  $(E_{1}^{\text{lab}}, E_{2}^{\text{lab}}, E_{3}^{\text{lab}}) = (358.7, 17.3, 26.8)\,\mathrm{GeV}$\newline
  $q_{\perp} = 7.28\,\mathrm{GeV}$, $k_{\perp} = 5.34\,\mathrm{GeV}$]
  {
    \label{fig:illustrateVectors_CMS:h}
    \includegraphics[width=0.48\linewidth]{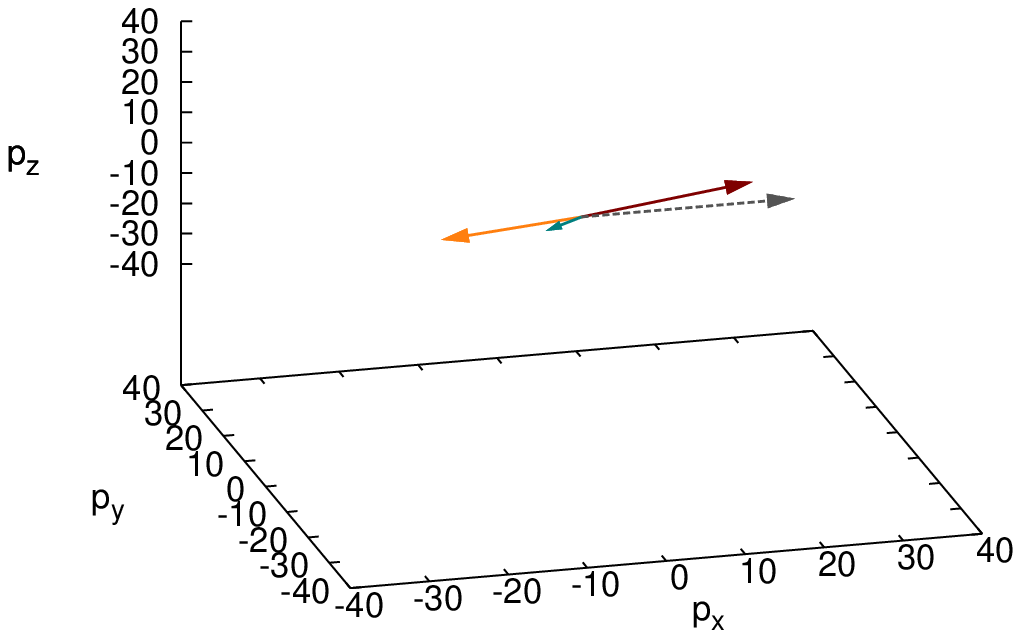}
  }

  \caption{(Color online) Same as \ref{fig:illustrateVectors_CMS1}. Part 2: Events 5-8 out of 8.}
  \label{fig:illustrateVectors_CMS2}
\end{figure}

In order to visualize the possible configurations of outgoing particles in $gg \rightarrow ggg$ events, the figures \ref{fig:illustrateVectors_CMS1} and \ref{fig:illustrateVectors_CMS2} illustrate a number of $gg \rightarrow ggg$ events in the center of momentum frame. These events feature an incoming $E=400\,\mathrm{GeV}$ gluon jet that interacts with constituents from a thermal medium ($T=400\,\mathrm{MeV}$). These events have been randomly selected obeying the relative importance given by the matrix element (\ref{eq:gg_to_ggg}). All events are rotated such that the incoming jet momentum in the center of momentum frame points along the positive x-direction and that the outgoing momentum $\vec{p}_1^{\,\prime}$ is in the x-y-plane.

\FloatBarrier

\bibliography{fochler}

\begin{thebibliography}{35}
\expandafter\ifx\csname natexlab\endcsname\relax\def\natexlab#1{#1}\fi
\expandafter\ifx\csname bibnamefont\endcsname\relax
  \def\bibnamefont#1{#1}\fi
\expandafter\ifx\csname bibfnamefont\endcsname\relax
  \def\bibfnamefont#1{#1}\fi
\expandafter\ifx\csname citenamefont\endcsname\relax
  \def\citenamefont#1{#1}\fi
\expandafter\ifx\csname url\endcsname\relax
  \def\url#1{\texttt{#1}}\fi
\expandafter\ifx\csname urlprefix\endcsname\relax\def\urlprefix{URL }\fi
\providecommand{\bibinfo}[2]{#2}
\providecommand{\eprint}[2][]{\url{#2}}

\bibitem[{\citenamefont{Adler et~al.}(2002)}]{Adler:2002xw}
\bibinfo{author}{\bibfnamefont{C.}~\bibnamefont{Adler}} \bibnamefont{et~al.}
  (\bibinfo{collaboration}{STAR}), \bibinfo{journal}{Phys. Rev. Lett.}
  \textbf{\bibinfo{volume}{89}}, \bibinfo{pages}{202301}
  (\bibinfo{year}{2002}), \eprint{nucl-ex/0206011}.

\bibitem[{\citenamefont{Adcox et~al.}(2002)}]{Adcox:2001jp}
\bibinfo{author}{\bibfnamefont{K.}~\bibnamefont{Adcox}} \bibnamefont{et~al.}
  (\bibinfo{collaboration}{PHENIX}), \bibinfo{journal}{Phys. Rev. Lett.}
  \textbf{\bibinfo{volume}{88}}, \bibinfo{pages}{022301}
  (\bibinfo{year}{2002}), \eprint{nucl-ex/0109003}.

\bibitem[{\citenamefont{Gyulassy and Wang}(1994)}]{Gyulassy:1993hr}
\bibinfo{author}{\bibfnamefont{M.}~\bibnamefont{Gyulassy}} \bibnamefont{and}
  \bibinfo{author}{\bibfnamefont{X.-N.} \bibnamefont{Wang}},
  \bibinfo{journal}{Nucl. Phys.} \textbf{\bibinfo{volume}{B420}},
  \bibinfo{pages}{583} (\bibinfo{year}{1994}), \eprint{nucl-th/9306003}.

\bibitem[{\citenamefont{Zakharov}(1996)}]{Zakharov:1996fv}
\bibinfo{author}{\bibfnamefont{B.~G.} \bibnamefont{Zakharov}},
  \bibinfo{journal}{JETP Lett.} \textbf{\bibinfo{volume}{63}},
  \bibinfo{pages}{952} (\bibinfo{year}{1996}), \eprint{hep-ph/9607440}.

\bibitem[{\citenamefont{Baier et~al.}(1997)\citenamefont{Baier, Dokshitzer,
  Mueller, Peigne, and Schiff}}]{Baier:1996sk}
\bibinfo{author}{\bibfnamefont{R.}~\bibnamefont{Baier}},
  \bibinfo{author}{\bibfnamefont{Y.~L.} \bibnamefont{Dokshitzer}},
  \bibinfo{author}{\bibfnamefont{A.~H.} \bibnamefont{Mueller}},
  \bibinfo{author}{\bibfnamefont{S.}~\bibnamefont{Peigne}}, \bibnamefont{and}
  \bibinfo{author}{\bibfnamefont{D.}~\bibnamefont{Schiff}},
  \bibinfo{journal}{Nucl. Phys.} \textbf{\bibinfo{volume}{B484}},
  \bibinfo{pages}{265} (\bibinfo{year}{1997}), \eprint{hep-ph/9608322}.

\bibitem[{\citenamefont{Baier et~al.}(1998)\citenamefont{Baier, Dokshitzer,
  Mueller, and Schiff}}]{Baier:1998yf}
\bibinfo{author}{\bibfnamefont{R.}~\bibnamefont{Baier}},
  \bibinfo{author}{\bibfnamefont{Y.~L.} \bibnamefont{Dokshitzer}},
  \bibinfo{author}{\bibfnamefont{A.~H.} \bibnamefont{Mueller}},
  \bibnamefont{and} \bibinfo{author}{\bibfnamefont{D.}~\bibnamefont{Schiff}},
  \bibinfo{journal}{Phys. Rev.} \textbf{\bibinfo{volume}{C58}},
  \bibinfo{pages}{1706} (\bibinfo{year}{1998}), \eprint{hep-ph/9803473}.

\bibitem[{\citenamefont{Gyulassy et~al.}(2001)\citenamefont{Gyulassy, Levai,
  and Vitev}}]{Gyulassy:2000er}
\bibinfo{author}{\bibfnamefont{M.}~\bibnamefont{Gyulassy}},
  \bibinfo{author}{\bibfnamefont{P.}~\bibnamefont{Levai}}, \bibnamefont{and}
  \bibinfo{author}{\bibfnamefont{I.}~\bibnamefont{Vitev}},
  \bibinfo{journal}{Nucl. Phys.} \textbf{\bibinfo{volume}{B594}},
  \bibinfo{pages}{371} (\bibinfo{year}{2001}), \eprint{nucl-th/0006010}.

\bibitem[{\citenamefont{Jeon and Moore}(2005)}]{Jeon:2003gi}
\bibinfo{author}{\bibfnamefont{S.}~\bibnamefont{Jeon}} \bibnamefont{and}
  \bibinfo{author}{\bibfnamefont{G.~D.} \bibnamefont{Moore}},
  \bibinfo{journal}{Phys. Rev.} \textbf{\bibinfo{volume}{C71}},
  \bibinfo{pages}{034901} (\bibinfo{year}{2005}), \eprint{hep-ph/0309332}.

\bibitem[{\citenamefont{Salgado and Wiedemann}(2003)}]{Salgado:2003gb}
\bibinfo{author}{\bibfnamefont{C.~A.} \bibnamefont{Salgado}} \bibnamefont{and}
  \bibinfo{author}{\bibfnamefont{U.~A.} \bibnamefont{Wiedemann}},
  \bibinfo{journal}{Phys. Rev.} \textbf{\bibinfo{volume}{D68}},
  \bibinfo{pages}{014008} (\bibinfo{year}{2003}), \eprint{hep-ph/0302184}.

\bibitem[{\citenamefont{Wicks et~al.}(2007)\citenamefont{Wicks, Horowitz,
  Djordjevic, and Gyulassy}}]{Wicks:2005gt}
\bibinfo{author}{\bibfnamefont{S.}~\bibnamefont{Wicks}},
  \bibinfo{author}{\bibfnamefont{W.}~\bibnamefont{Horowitz}},
  \bibinfo{author}{\bibfnamefont{M.}~\bibnamefont{Djordjevic}},
  \bibnamefont{and} \bibinfo{author}{\bibfnamefont{M.}~\bibnamefont{Gyulassy}},
  \bibinfo{journal}{Nucl. Phys.} \textbf{\bibinfo{volume}{A784}},
  \bibinfo{pages}{426} (\bibinfo{year}{2007}), \eprint{nucl-th/0512076}.

\bibitem[{\citenamefont{Romatschke and Romatschke}(2007)}]{Romatschke:2007mq}
\bibinfo{author}{\bibfnamefont{P.}~\bibnamefont{Romatschke}} \bibnamefont{and}
  \bibinfo{author}{\bibfnamefont{U.}~\bibnamefont{Romatschke}},
  \bibinfo{journal}{Phys. Rev. Lett.} \textbf{\bibinfo{volume}{99}},
  \bibinfo{pages}{172301} (\bibinfo{year}{2007}), \eprint{0706.1522}.

\bibitem[{\citenamefont{Kovtun et~al.}(2005)\citenamefont{Kovtun, Son, and
  Starinets}}]{Kovtun:2004de}
\bibinfo{author}{\bibfnamefont{P.}~\bibnamefont{Kovtun}},
  \bibinfo{author}{\bibfnamefont{D.~T.} \bibnamefont{Son}}, \bibnamefont{and}
  \bibinfo{author}{\bibfnamefont{A.~O.} \bibnamefont{Starinets}},
  \bibinfo{journal}{Phys. Rev. Lett.} \textbf{\bibinfo{volume}{94}},
  \bibinfo{pages}{111601} (\bibinfo{year}{2005}), \eprint{hep-th/0405231}.

\bibitem[{\citenamefont{Bass et~al.}(2009)}]{Bass:2008rv}
\bibinfo{author}{\bibfnamefont{S.~A.} \bibnamefont{Bass}} \bibnamefont{et~al.},
  \bibinfo{journal}{Phys. Rev.} \textbf{\bibinfo{volume}{C79}},
  \bibinfo{pages}{024901} (\bibinfo{year}{2009}), \eprint{0808.0908}.

\bibitem[{\citenamefont{Schenke et~al.}(2009)\citenamefont{Schenke, Gale, and
  Jeon}}]{Schenke:2009gb}
\bibinfo{author}{\bibfnamefont{B.}~\bibnamefont{Schenke}},
  \bibinfo{author}{\bibfnamefont{C.}~\bibnamefont{Gale}}, \bibnamefont{and}
  \bibinfo{author}{\bibfnamefont{S.}~\bibnamefont{Jeon}},
  \bibinfo{journal}{Phys. Rev.} \textbf{\bibinfo{volume}{C80}},
  \bibinfo{pages}{054913} (\bibinfo{year}{2009}), \eprint{0909.2037}.

\bibitem[{\citenamefont{Fochler et~al.}(2009)\citenamefont{Fochler, Xu, and
  Greiner}}]{Fochler:2008ts}
\bibinfo{author}{\bibfnamefont{O.}~\bibnamefont{Fochler}},
  \bibinfo{author}{\bibfnamefont{Z.}~\bibnamefont{Xu}}, \bibnamefont{and}
  \bibinfo{author}{\bibfnamefont{C.}~\bibnamefont{Greiner}},
  \bibinfo{journal}{Phys. Rev. Lett.} \textbf{\bibinfo{volume}{102}},
  \bibinfo{pages}{202301} (\bibinfo{year}{2009}), \eprint{0806.1169}.

\bibitem[{\citenamefont{Xu and Greiner}(2005)}]{Xu:2004mz}
\bibinfo{author}{\bibfnamefont{Z.}~\bibnamefont{Xu}} \bibnamefont{and}
  \bibinfo{author}{\bibfnamefont{C.}~\bibnamefont{Greiner}},
  \bibinfo{journal}{Phys. Rev.} \textbf{\bibinfo{volume}{C71}},
  \bibinfo{pages}{064901} (\bibinfo{year}{2005}), \eprint{hep-ph/0406278}.

\bibitem[{\citenamefont{Xu and Greiner}(2007)}]{Xu:2007aa}
\bibinfo{author}{\bibfnamefont{Z.}~\bibnamefont{Xu}} \bibnamefont{and}
  \bibinfo{author}{\bibfnamefont{C.}~\bibnamefont{Greiner}},
  \bibinfo{journal}{Phys. Rev.} \textbf{\bibinfo{volume}{C76}},
  \bibinfo{pages}{024911} (\bibinfo{year}{2007}), \eprint{hep-ph/0703233}.

\bibitem[{\citenamefont{Gunion and Bertsch}(1982)}]{Gunion:1981qs}
\bibinfo{author}{\bibfnamefont{J.~F.} \bibnamefont{Gunion}} \bibnamefont{and}
  \bibinfo{author}{\bibfnamefont{G.}~\bibnamefont{Bertsch}},
  \bibinfo{journal}{Phys. Rev.} \textbf{\bibinfo{volume}{D25}},
  \bibinfo{pages}{746} (\bibinfo{year}{1982}).

\bibitem[{\citenamefont{Migdal}(1956)}]{Migdal:1956tc}
\bibinfo{author}{\bibfnamefont{A.~B.} \bibnamefont{Migdal}},
  \bibinfo{journal}{Phys. Rev.} \textbf{\bibinfo{volume}{103}},
  \bibinfo{pages}{1811} (\bibinfo{year}{1956}).

\bibitem[{\citenamefont{Zapp et~al.}(2009)\citenamefont{Zapp, Stachel, and
  Wiedemann}}]{Zapp:2008af}
\bibinfo{author}{\bibfnamefont{K.}~\bibnamefont{Zapp}},
  \bibinfo{author}{\bibfnamefont{J.}~\bibnamefont{Stachel}}, \bibnamefont{and}
  \bibinfo{author}{\bibfnamefont{U.~A.} \bibnamefont{Wiedemann}},
  \bibinfo{journal}{Phys. Rev. Lett.} \textbf{\bibinfo{volume}{103}},
  \bibinfo{pages}{152302} (\bibinfo{year}{2009}), \eprint{0812.3888}.

\bibitem[{\citenamefont{Qin et~al.}(2008)}]{Qin:2007rn}
\bibinfo{author}{\bibfnamefont{G.-Y.} \bibnamefont{Qin}} \bibnamefont{et~al.},
  \bibinfo{journal}{Phys. Rev. Lett.} \textbf{\bibinfo{volume}{100}},
  \bibinfo{pages}{072301} (\bibinfo{year}{2008}), \eprint{0710.0605}.

\bibitem[{\citenamefont{Majumder et~al.}(2007)\citenamefont{Majumder, Nonaka,
  and Bass}}]{Majumder:2007ae}
\bibinfo{author}{\bibfnamefont{A.}~\bibnamefont{Majumder}},
  \bibinfo{author}{\bibfnamefont{C.}~\bibnamefont{Nonaka}}, \bibnamefont{and}
  \bibinfo{author}{\bibfnamefont{S.~A.} \bibnamefont{Bass}},
  \bibinfo{journal}{Phys. Rev.} \textbf{\bibinfo{volume}{C76}},
  \bibinfo{pages}{041902} (\bibinfo{year}{2007}), \eprint{nucl-th/0703019}.

\bibitem[{\citenamefont{Qin et~al.}(2007)}]{Qin:2007zz}
\bibinfo{author}{\bibfnamefont{G.-Y.} \bibnamefont{Qin}} \bibnamefont{et~al.},
  \bibinfo{journal}{Phys. Rev.} \textbf{\bibinfo{volume}{C76}},
  \bibinfo{pages}{064907} (\bibinfo{year}{2007}), \eprint{0705.2575}.

\bibitem[{\citenamefont{Renk et~al.}(2007)\citenamefont{Renk, Ruppert, Nonaka,
  and Bass}}]{Renk:2006sx}
\bibinfo{author}{\bibfnamefont{T.}~\bibnamefont{Renk}},
  \bibinfo{author}{\bibfnamefont{J.}~\bibnamefont{Ruppert}},
  \bibinfo{author}{\bibfnamefont{C.}~\bibnamefont{Nonaka}}, \bibnamefont{and}
  \bibinfo{author}{\bibfnamefont{S.~A.} \bibnamefont{Bass}},
  \bibinfo{journal}{Phys. Rev.} \textbf{\bibinfo{volume}{C75}},
  \bibinfo{pages}{031902} (\bibinfo{year}{2007}), \eprint{nucl-th/0611027}.

\bibitem[{\citenamefont{Chen et~al.}(2010)\citenamefont{Chen, Greiner, Wang,
  Wang, and Xu}}]{Chen:2010te}
\bibinfo{author}{\bibfnamefont{X.-F.} \bibnamefont{Chen}},
  \bibinfo{author}{\bibfnamefont{C.}~\bibnamefont{Greiner}},
  \bibinfo{author}{\bibfnamefont{E.}~\bibnamefont{Wang}},
  \bibinfo{author}{\bibfnamefont{X.-N.} \bibnamefont{Wang}}, \bibnamefont{and}
  \bibinfo{author}{\bibfnamefont{Z.}~\bibnamefont{Xu}} (\bibinfo{year}{2010}),
  \eprint{1002.1165}.

\bibitem[{\citenamefont{Xu et~al.}(2008)\citenamefont{Xu, Greiner, and
  Stocker}}]{Xu:2007jv}
\bibinfo{author}{\bibfnamefont{Z.}~\bibnamefont{Xu}},
  \bibinfo{author}{\bibfnamefont{C.}~\bibnamefont{Greiner}}, \bibnamefont{and}
  \bibinfo{author}{\bibfnamefont{H.}~\bibnamefont{Stocker}},
  \bibinfo{journal}{Phys. Rev. Lett.} \textbf{\bibinfo{volume}{101}},
  \bibinfo{pages}{082302} (\bibinfo{year}{2008}), \eprint{0711.0961}.

\bibitem[{\citenamefont{Xu and Greiner}(2009)}]{Xu:2008av}
\bibinfo{author}{\bibfnamefont{Z.}~\bibnamefont{Xu}} \bibnamefont{and}
  \bibinfo{author}{\bibfnamefont{C.}~\bibnamefont{Greiner}},
  \bibinfo{journal}{Phys. Rev.} \textbf{\bibinfo{volume}{C79}},
  \bibinfo{pages}{014904} (\bibinfo{year}{2009}), \eprint{0811.2940}.

\bibitem[{\citenamefont{Xu and Greiner}(2008)}]{Xu:2007ns}
\bibinfo{author}{\bibfnamefont{Z.}~\bibnamefont{Xu}} \bibnamefont{and}
  \bibinfo{author}{\bibfnamefont{C.}~\bibnamefont{Greiner}},
  \bibinfo{journal}{Phys. Rev. Lett.} \textbf{\bibinfo{volume}{100}},
  \bibinfo{pages}{172301} (\bibinfo{year}{2008}), \eprint{0710.5719}.

\bibitem[{\citenamefont{Li et~al.}(2009)}]{Li:2009ti}
\bibinfo{author}{\bibfnamefont{W.}~\bibnamefont{Li}} \bibnamefont{et~al.},
  \bibinfo{journal}{Phys. Rev.} \textbf{\bibinfo{volume}{C80}},
  \bibinfo{pages}{064913} (\bibinfo{year}{2009}), \eprint{0903.2165}.

\bibitem[{\citenamefont{Renk}(2008)}]{Renk:2008xq}
\bibinfo{author}{\bibfnamefont{T.}~\bibnamefont{Renk}}, \bibinfo{journal}{Phys.
  Rev.} \textbf{\bibinfo{volume}{C78}}, \bibinfo{pages}{034904}
  (\bibinfo{year}{2008}), \eprint{0803.0218}.

\bibitem[{\citenamefont{Adare et~al.}(2008)}]{Adare:2008qa}
\bibinfo{author}{\bibfnamefont{A.}~\bibnamefont{Adare}} \bibnamefont{et~al.}
  (\bibinfo{collaboration}{PHENIX}), \bibinfo{journal}{Phys. Rev. Lett.}
  \textbf{\bibinfo{volume}{101}}, \bibinfo{pages}{232301}
  (\bibinfo{year}{2008}), \eprint{0801.4020}.

\bibitem[{\citenamefont{Gluck et~al.}(1995)\citenamefont{Gluck, Reya, and
  Vogt}}]{Gluck:1994uf}
\bibinfo{author}{\bibfnamefont{M.}~\bibnamefont{Gluck}},
  \bibinfo{author}{\bibfnamefont{E.}~\bibnamefont{Reya}}, \bibnamefont{and}
  \bibinfo{author}{\bibfnamefont{A.}~\bibnamefont{Vogt}}, \bibinfo{journal}{Z.
  Phys.} \textbf{\bibinfo{volume}{C67}}, \bibinfo{pages}{433}
  (\bibinfo{year}{1995}).

\bibitem[{\citenamefont{Abelev et~al.}(2008)}]{Abelev:2008ed}
\bibinfo{author}{\bibfnamefont{B.~I.} \bibnamefont{Abelev}}
  \bibnamefont{et~al.} (\bibinfo{collaboration}{STAR}), \bibinfo{journal}{Phys.
  Rev.} \textbf{\bibinfo{volume}{C77}}, \bibinfo{pages}{054901}
  (\bibinfo{year}{2008}), \eprint{0801.3466}.

\bibitem[{\citenamefont{Gossiaux et~al.}(2010)\citenamefont{Gossiaux, Aichelin,
  Gousset, and Guiho}}]{Gossiaux:2010yx}
\bibinfo{author}{\bibfnamefont{P.~B.} \bibnamefont{Gossiaux}},
  \bibinfo{author}{\bibfnamefont{J.}~\bibnamefont{Aichelin}},
  \bibinfo{author}{\bibfnamefont{T.}~\bibnamefont{Gousset}}, \bibnamefont{and}
  \bibinfo{author}{\bibfnamefont{V.}~\bibnamefont{Guiho}}
  (\bibinfo{year}{2010}), \eprint{1001.4166}.

\bibitem[{\citenamefont{Bouras et~al.}(2009)}]{Bouras:2009nn}
\bibinfo{author}{\bibfnamefont{I.}~\bibnamefont{Bouras}} \bibnamefont{et~al.},
  \bibinfo{journal}{Phys. Rev. Lett.} \textbf{\bibinfo{volume}{103}},
  \bibinfo{pages}{032301} (\bibinfo{year}{2009}), \eprint{0902.1927}.

\end{thebibliography}

\end{document}